\documentclass[prd,
              preprint,
              superscriptaddress,
              preprintnumbers,
              eqsecnum,
              showpacs,
              nofootinbib,
              nobibnotes,
              noeprint,
              nolongbibliography
              ]{revtex4-2}

\usepackage[linkcolor=blue,
            citecolor=blue,
            urlcolor=blue,
            colorlinks=true,
            breaklinks
            ]{hyperref}

\usepackage{booktabs}
\usepackage[italic]{hepnames}
\usepackage{amsfonts}
\usepackage{amssymb}
\usepackage{amsthm}
\usepackage{bm}
\usepackage{slashed}
\usepackage{dsfont}
\usepackage{tikz}
\usepackage[dvipsnames]{xcolor}
\usepackage[shortlabels]{enumitem}
\usepackage{xspace}
\usepackage{siunitx}
\usepackage{bookmark}
\usepackage{bbm}

\def\1eq#1{Eq.\nobreak\thinspace(\ref{#1})}

\def\2eqs#1#2{Eqs.\nobreak\thinspace(\ref{#1}) and\nobreak\thinspace(\ref{#2})}
\def\3eqs#1#2#3{Eqs.\nobreak\thinspace(\ref{#1}),\nobreak\thinspace(\ref{#2}) and\nobreak\thinspace(\ref{#3})}

\def\fig#1{\hyperref[#1]{Fig.\nobreak\thinspace\ref*{#1}}}
\def\figA#1{\hyperref[#1]{Fig.\nobreak\thinspace\ref*{#1}A}}
\def\figB#1{\hyperref[#1]{Fig.\nobreak\thinspace\ref*{#1}B}}
\def\figC#1{\hyperref[#1]{Fig.\nobreak\thinspace\ref*{#1}C}}
\def\figs2#1#2{\hyperref[#1]{Figs.\nobreak\thinspace\ref*{#1}}\nobreak\thinspace and\nobreak\thinspace\hyperref[#2]{\ref*{#2}}}
\def\figs3#1#2#3{\hyperref[#1]{Figs.\nobreak\thinspace\ref*{#1}},\nobreak\thinspace\hyperref[#2]{\ref*{#2}}\nobreak\thinspace and\nobreak\thinspace\hyperref[#3]{\ref*{#3}}}

\def\tab#1{\hyperref[#1]{Tab.\nobreak\thinspace\ref*{#1}}}

\def\sect#1{\hyperref[#1]{Sec.\nobreak\thinspace\ref*{#1}}}
\def\appref#1{\hyperref[#1]{App.\nobreak\thinspace\ref*{#1}}}

\def\ie{{\it i.e.}, }
\def\eg{{\it e.g.}, }

\newcommand{\be}{\begin{equation}}
\newcommand{\ee}{\end{equation}}

\newcommand{\bea}{\begin{eqnarray}}
\newcommand{\eea}{\end{eqnarray}}

\def\g{\Gamma}              
\def\gt{\overline{\Gamma}}  

\def\s#1{{\scriptscriptstyle #1}}

\def\MOMt{$\widetilde{\text{MOM}}$}

\def\srm#1{{\rm{\scriptscriptstyle #1}}}

\newcommand{\Ls}{ \mathit{L}_{{sg}}}

\def\kvec{\mathbf{k}}

\begin{document}

\title{\boldmath  Quark-gluon vertex in the complex plane}

\author{M.N.~Ferreira}
\email{narciso.ferreira@ufrgs.br}
\affiliation{\mbox{Instituto de Física, Universidade Federal do Rio Grande do Sul}, Caixa Postal 15051, 91501-970, Porto Alegre, RS, Brazil}

\author{A.S.~Miramontes}
\email{angel.s.miramontes@uv.es}
\affiliation{\mbox{Department of Theoretical Physics and IFIC, University of Valencia and CSIC}, E-46100, Valencia, Spain}

\author{J.M.~Morgado}
\email{jose.m.morgado@uv.es}
\affiliation{\mbox{Department of Theoretical Physics and IFIC, University of Valencia and CSIC}, E-46100, Valencia, Spain}

\author{J.~Papavassiliou}
\email{joannis.papavassiliou@uv.es}
\affiliation{\mbox{Department of Theoretical Physics and IFIC, University of Valencia and CSIC}, E-46100, Valencia, Spain}

\begin{abstract}
In the present work we explore
for the first time the general structure and properties of the nonperturbative quark-gluon vertex in the complex plane. 
Specifically, we focus on  
the transversely-projected quark-gluon vertex that emerges from a recently developed symmetry-preserving approach for  the study of  
meson properties beyond the rainbow-ladder approximation.  
The analysis focuses on the so-called ``soft-gluon" limit, which 
reduces the momentum-dependence of the 
corresponding vertex form factors 
to a single momentum variable.
The complexification of this variable
inside the defining integrals 
furnishes unambiguously 
all eight vertex form factors 
within a concrete domain of the complex variable, delimited by a characteristic parabola. 
The extent of this reliable domain is 
determined by the appearance of the first singularity in the integrands of the vertex integrals, where the standard Wick rotation must be duly supplemented by additional crucial contributions.  
This primary analytic region 
may be extended considerably 
by resorting to standard extrapolation methods, which remain valid 
up until the appearance of  
complex structures
associated with the onset of physical 
processes. The generalization of the method to arbitrary gluon momenta, and 
its relevance for the determination of  
the quark propagator in the complex plane, are briefly discussed. 

\end{abstract}

\maketitle

\newpage 


\section{Introduction}\label{sec:intro}

In recent years,  
the nonperturbative structure and properties of the main QCD correlation (Green's) functions 
have been rather successfully 
determined, for space-like (Euclidean) momenta, through the synergy of 
functional methods and gauge-fixed lattice simulations~\cite{Alkofer:2000wg,Pawlowski:2003hq,Binosi:2009qm,Maas:2011se,Cloet:2013jya,Huber:2018ned,Dupuis:2020fhh,Ferreira:2023fva,Ferreira:2025anh,Huber:2025cbd}. Instead, the extension 
of these functions 
to the time-like or complex domain, 
required for carrying out real-time computations, is still far from complete; for related works see, \eg  \cite{Alkofer:2003jj,Strauss:2012dg,Frederico:2013vga, Dudal:2013yva,dePaula:2016oct, El-Bennich:2016qmb,Tripolt:2018xeo,Binosi:2019ecz,Dudal:2019gvn,Eichmann:2019dts,Li:2019hyv,Li:2021wol,Fischer:2020xnb,Horak:2021pfr,Horak:2021syv,Horak:2022myj,Eichmann:2021vnj,Horak:2022aza,Huber:2022nzs,Huber:2023uzd,Pawlowski:2024kxc}. 

Particularly important in these considerations is the quark-gluon vertex, $\g^\mu(q,r,-p)$.
Given the central role of this vertex in contemporary hadron physics, its 
nonperturbative aspects 
have been studied in Euclidean space 
with continuous 
functional approaches \cite{Bhagwat:2004kj,LlanesEstrada:2004jz,Matevosyan:2006bk,Fischer:2006ub,Aguilar:2018epe,Qin:2013mta,Hopfer:2012cnq,Rojas:2013tza,Williams:2014iea,Williams:2015cvx,Sanchis-Alepuz:2015qra,Pelaez:2015tba,Alkofer:2008tt,Mitter:2014wpa,Cyrol:2017ewj,Aguilar:2014lha,Gao:2021wun,Aguilar:2016lbe,Oliveira:2018ukh,Albino:2018ncl,Tang:2019zbk,Huber:2018ned,Albino:2021rvj,Windisch:2012de, Oliveira:2020yac, Wieland:2026iml}, and through a 
multitude of lattice simulations~\cite{Skullerud:2002sk,Skullerud:2002ge,Skullerud:2003qu,Skullerud:2004gp,Lin:2005zd,Kizilersu:2021jen,Kizilersu:2006et,Sternbeck:2017ntv,Skullerud:2021pel,Oliveira:2016muq,Oliveira:2018fkj}.
However, with the exception of the general analysis presented 
in~\cite{Huber:2022nzs,Huber:2023uzd}, 
the properties of the quark-gluon vertex in the complex plane remain largely unexplored. 

Attaining a quantitative understanding of the evolution of the quark-gluon vertex for complex momenta becomes all the more timely in light of recent developments in the physics of mesons.
In particular, a symmetry-preserving approach has been developed in a series of works \cite{Miramontes:2025imd,Ferreira:2025wpu,Ferreira:2026gbe}, where fully-dressed quark-gluon vertices are self-consistently included in the dynamical equations that determine 
the meson spectrum. Specifically, the  
presence of the quark-gluon vertex in the meson Bethe-Salpeter equation (BSE) \cite{Salpeter:1951sz,PhysRev.84.350,Bethe1957,Nakanishi:1969ph,Jain:1993qh,Munczek:1994zz}
requires knowledge of its form factors in the complex plane, 
given that
the mass condition $P^2=-M^2$, with 
\mbox{$P=(0,0,0, iM)$},  introduces 
complex momenta in the corresponding integrals. 
This situation is to be contrasted with 
the 
standard rainbow-ladder
approximation, where, 
due to the simplification 
$\g^\mu(q,r,-p) \to \gamma^\mu$, 
only the complex structure of the quark propagator is required, see, \eg \cite{Fischer:2005en,Windisch:2016iud, Sanchis-Alepuz:2017jjd, Huber:2025cbd}. 

In the present work we report on an exploratory study of the  
transversely-projected quark-gluon vertex (Landau gauge)
in the complex plane. 
Specifically, we consider the 
vertex that arises after the 
specific approximation 
introduced  in \cite{Ferreira:2025wpu,Ferreira:2026gbe} has been implemented. 
In particular, the ``one-loop dressed" version of 
the Schwinger-Dyson equation (SDE) that governs the 
vertex $\g^\mu(q,r,-p)$ is composed by two Feynman diagrams 
with fully dressed vertices and propagators~\cite{Alkofer:2008tt,Williams:2015cvx,Aguilar:2024ciu}. The approximation of 
\cite{Ferreira:2025wpu,Ferreira:2026gbe} replaces all internal 
quark-gluon vertices, generically denoted by 
$\g^\mu(q,r,-p)$, by 
\mbox{$\g^\mu(q,r,-p) \to V(q^2) 
\gamma^\mu$}
where $q$ denotes the momentum of the gluon.   
The function 
$V(q^2)$ corresponds to the 
form factor associated with the 
classical tensorial structure, 
evaluated   
in the so-called 
``symmetric'' kinematic configuration, 
defined as \mbox{$q^2=r^2=p^2$}. 
Thus, the resulting SDE 
simplifies to the one shown in \fig{fig:greenqgsde}. 
We emphasize that the resulting 
$\g^\mu(q,r,-p)$ (cyan vertex in 
\fig{fig:greenqgsde}) 
displays the full kinematic 
dependence associated with a quark-gluon vertex, namely 
it is composed by eight 
tensorial structures, multiplied by the attendant form factors
$\lambda_i(q,r,-p)$. 
In addition, and quite importantly, within this approximation the $\lambda_i(q,r,-p)$ are obtained 
by simple integration of 
appropriately projected integral expressions, \ie no 
iterative procedure is required. 

In order to facilitate the analysis, 
we restrict our study to 
the simple kinematic case   
known as ``{\it soft-gluon}" kinematics,  
where the external gluon momentum $q$ vanishes. 
We stress that this limit is 
not implemented on the 
tensorial basis, 
which retains its full kinematic structure, but rather at the 
level of the corresponding form factors. 
Thus, one obtains the eight basis tensors with their full kinematic dependence, while 
the associated form factors 
are functions of a single momentum, 
$\lambda_i(0,p,-p) := \lambda^{sg}_i(p^2)$.
Then, the study focuses on the 
behavior of these  
$\lambda^{sg}_i(p^2)$
as one complexifies the momentum, \ie sets $p^2 \to z$, with $z \in  \mathbb{C}$, at the level of the integral expressions
that define the $\lambda^{sg}_i(p^2)$. 

The treatment is further simplified 
by employing 
a particular {\it Ansatz} for the 
quark propagator entering in the 
aforementioned integrals, instead of  
obtaining it dynamically from the corresponding gap equation. Specifically, we opt for 
a quark-propagator that exhibits a pair 
of complex conjugate poles, as reported 
in a variety of studies \cite{Alkofer:2003jj,Fischer:2008sp,Dorkin:2013rsa,Windisch:2016iud,Eichmann:2016yit, Miramontes:2019mco,Gao:2024gdj,Alkofer:2026vux}. To be sure, this 
contentious feature has been associated with 
an excessive vertex strength in the quark gap equation \cite{Pawlowski:2024kxc},  
and is absent from   
the recent comprehensive study of \cite{Wieland:2026iml}; 
nonetheless, we adopt it 
as a reasonable operating hypothesis for 
the present investigation.

Since, in the soft-gluon limit, 
the single external momentum 
$p$ may be channeled 
entirely through the 
quark propagators inside the vertex Feynman diagrams,
all other components, 
\ie  the gluon propagators,  
the form-factor of the 
three-gluon vertex, and the function $V(q^2)$, 
are evaluated for Euclidean momenta. 
As a result, 
the integrals defining the $\lambda^{sg}_i(z)$ may be carried out directly 
within a certain domain of the $z$,
yielding unambiguously 
the real and imaginary parts of 
$\lambda^{sg}_i(z)$.
The extent of this domain is delimited 
by the points where 
the quark-propagator inside the integrals 
becomes singular. Then, in order to compute further,  
the inclusion of the residue of the integrand
at that singular point 
is needed; however, this 
requires the knowledge of all aforementioned 
components not only on the 
Euclidean axis but in the  
complex plane, a fact that 
prevents the continuation of the calculation.

Nonetheless, the range of 
the computations may be considerably 
extended by resorting to 
the Schlessinger point method (SPM) \cite{Schlessinger:1968vsk, Tripolt:2018xeo, Binosi:2019ecz}. This method replaces the aforementioned computation of the residue,
and prolongs the $\lambda^{sg}_i(z)$
until the appearance of the branch cut associated with the physical thresholds. Specifically, we construct SPM approximants that interpolate the $\lambda^{sg}_i(p^2)$ obtained by direct integration for real $p^2$, within the range of validity of the standard Euclidean integral. Numerical analytic continuation is then performed by complexifying the argument $p^2$ of the approximants.  The resulting SPM reconstructions agree with the $\lambda^{sg}_i(p^2)$ within the domain where they can be computed directly, and are expected to accurately predict the complex $\lambda^{sg}_i(z)$ until the position of their nearest Landau singularities
\cite{Landau:1959fi,Eden:1966dnq,Itzykson:1980rh}. 

The article is organized as follows. In \sect{sec:qgsde} we review the key features 
of the transversely-projected quark-gluon vertex that we will consider in this study. 
 Then, in \sect{sec:sglim}, we introduce the soft-gluon kinematic limit of this vertex, and 
 present the dynamical equations for all of its form factors. In \sect{sec:anstruct} we 
explore in detail the analytic structure of 
a typical one-loop vertex integral, which serves as a   
benchmark of the ensuing nonperturbative 
treatment. 
 In \sect{sec:num},  we present the direct numerical evaluation of the corresponding form factors in the complex plane, and their extrapolation beyond the formally accessible region by means of the SPM.
 In \sect{sec:conc} we summarize our main findings and conclusions. Finally, in \appref{app:coefs} we list the defining coefficients for the SPM extrapolations 
 employed in this work.

\section{The quark-gluon vertex}\label{sec:qgsde}

In this section we briefly  
review the main properties 
of the transversely-projected quark-gluon vertex  associated with the 
symmetry-preserving treatment of \cite{Ferreira:2025wpu,Ferreira:2026gbe}. 

Throughout this work we employ the 
standard Landau gauge, where the 
gluon propagator,
$\Delta_{\mu\nu}^{ab}(q)=-i\delta^{ab}\Delta_{\mu\nu}(q)$,
assumes 
a completely transverse form 
\begin{align}
\Delta_{\mu\nu}(q)&= P_{\mu\nu}(q)\Delta(q^2)\,,& P_{\mu\nu}(q)&=g_{\mu\nu}-\frac{q_\mu q_\nu}{q^2}\,.
\end{align}
In addition, we denote by \mbox{$S^{ab}(p)=i\delta^{ab}S(p)$} the quark propagator, see \eg \cite{Itzykson:1980rh}; the standard decomposition
of its inverse is given by
\begin{align} 
\label{qprop}
    S^{-1}(p)=A(p^2)\slashed{p}-B(p^2)\,,
\end{align}
where $A(p^2)$ and $B(p^2)$ are the Dirac vector and scalar components, respectively. 
Note that the constituent quark mass function, 
${\mathcal M}(p^2)$,  
is defined as \mbox{${\mathcal M}(p^2) = B(p^2)/ A(p^2)$}.

We next consider the quark-gluon vertex, 
$\Gamma^a_{\!\mu}(q,r,-p)=igt^a\Gamma_{\!\mu}(q,r,-p)$, 
with $t^a$ the generators of the SU($N$) color group in the fundamental representation and $g$ the gauge coupling. 
In the Landau gauge, it is natural to consider  
the transversely projected quark-gluon vertex, $\overline\Gamma_{\!\mu}(q,r,-p)$,
and its tree-level counterpart, 
$\gt_{\!0}^\mu(q)$, 
defined as 
\be
\gt^\mu(q,r,-p) = 
P^{\mu\nu}(q) \Gamma_{\!\nu}(q,r,-p)\,, \qquad\qquad
\gt_{\!0}^\mu(q) 
=P^{\mu\nu}(q) \gamma_{\nu} \,.
\label{Gtrans}
\ee

In general kinematics, $\overline{\Gamma}_{\!\mu}(q,r,-p)$ is composed by eight independent tensors, $\bar\tau_{i}$, 
namely  
\begin{align} 
\label{decomp}
    \overline{\Gamma}^\mu(q,r,-p)=\sum_{i=1}^{8}\lambda_i(q,r,-p)\bar\tau_{i}^\mu(r,-p) \,, 
    \qquad\qquad  
 \bar\tau_{i}^\mu(r,-p)=  P^\mu_{\nu}(q)\tau_{i}^\nu(r,-p) \,,
\end{align}
where the closed expressions 
of the $\tau_{i}^\nu(r,-p)$ 
are given in Eq.~(2.9) of 
\cite{Aguilar:2024ciu}.

As was shown in detail in 
a series of recent articles \cite{Miramontes:2025imd,Ferreira:2025wpu,Ferreira:2026gbe},
one may employ 
fully-dressed quark-gluon vertices in the 
description of mesons dynamics, while  
maintaining intact the crucial 
Ward-Takahashi identities \cite{Itzykson:1980rh, Miransky:1994vk} associated with the chiral symmetry. The general findings of \cite{Ferreira:2025wpu} have been streamlined in \cite{Ferreira:2026gbe}, leading to  
a special form of the 
dynamical equation
satisfied by the quark-gluon vertex. 
Specifically, starting 
with the standard form 
of the SDE for the 
quark-gluon vertex \cite{Alkofer:2008tt,Williams:2015cvx,Aguilar:2024ciu}, 
see \eg Fig.~1A.~in~\cite{Ferreira:2026gbe}, 
we carry out inside the Feynman diagrams the 
substitution  
\be\
\label{eq:qgsym}
\g_\mu(q,r,-p) \to V_\mu(q)\,, \qquad \qquad V_\mu(q)= V(q^2)\gamma_\mu=\hspace{-4cm}\raisebox{-1cm}{\includegraphics[scale=0.85]{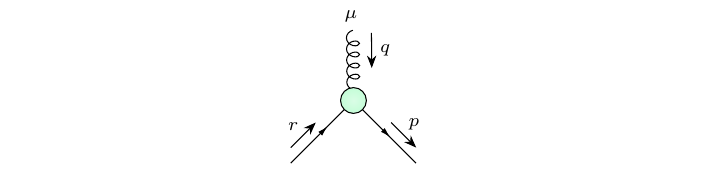}}\hspace{-4cm}\,.
\ee
Evidently, the $q$ 
in \1eq{eq:qgsym} represents a 
generic gluon momentum entering into a quark-gluon vertex; thus, 
in diagram ($c_1$) we have 
three different cases,  
$q \to q, k, k$, 
while in $(c_2)$
we have two cases, $q \to k, k-q$.

The function 
$V(q^2)$ is determined as follows. 
First, the SDE of the quark-gluon vertex \cite{Alkofer:2008tt,Blum:2014gna,Williams:2015cvx,Aguilar:2024ciu} derived in the three-particle-irreducible formalism \cite{Cornwall:1973ts,Cornwall:1974vz,Berges:2004pu,Carrington:2010qq,Williams:2015cvx}
is solved iteratively, maintaining the 
full momentum-dependence of the vertices inside the 
diagrams. Then, the form factor $\widetilde\lambda_1(q,r,-p)$,
associated with the tree-level (classical) tensor 
$\gamma_{\mu}$, is considered, and 
its slice corresponding to the 
symmetric configuration, $q^2=r^2=p^2$,  is singled out,
and identified with $V(q^2)$, 
\ie 
$V(q^2) = \widetilde{\lambda}_1 (q^2,q^2,\pi/3)$, see lower-left panel in \fig{fig:input}.  
Due to this particular feature, this approach was coined ``symmetric-vertex" 
approximation \cite{Ferreira:2026gbe}.

\begin{figure}[!t]
\hspace*{-1.25cm}
\includegraphics[scale=1.1]{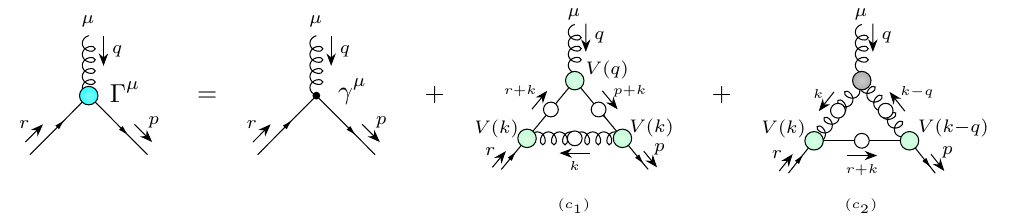}
\caption{The SDE of the quark-gluon vertex (cyan circle), defined by the substitution $\g_\mu\to V_\mu$ (green circles) in its standard one-loop dressed representation \cite{Alkofer:2008tt,Blum:2014gna,Williams:2015cvx,Aguilar:2024ciu} .}
\label{fig:greenqgsde}
\end{figure}

Note that even though 
the various $V_\mu(q)$ 
entering in the graphs on the 
r.h.s of 
\fig{fig:greenqgsde}
contain only the classical tensor, 
the result  
possesses the {\it full} kinematic 
structure associated with a quark-gluon vertex. In particular, as dictated by \1eq{decomp}, 
the resulting vertex $\g^\mu(q,r,-p)$ (cyan circle in 
\fig{fig:greenqgsde}) 
is composed by eight tensorial structures, whose 
 form factors, $\lambda_i$, depend on three kinematic variables, \eg $r^2$, $p^2$, and $\theta_{rp}$. 
 In addition, since  the $V(q^2)$ are supplied to $(c_1)$ and 
 $(c_2)$ as external input, 
 the form factors 
 $\lambda_i$ of the cyan vertex 
 are obtained through direct integration, rather than iteration.

With these approximations, the quark-gluon vertex SDE reads
\be\label{eq:vertexsde}
\overline{\Gamma}^\mu(q,r,-p)=\overline{\Gamma}^\mu_{\!0}(q)+c_1^\mu(q,r-p)+c_2^\mu(q,r,-p)\,,
\ee
with
\bea\label{eq:qgsdeTgreenAB}
\displaystyle c_1^\mu  & \displaystyle = & \displaystyle ic_a\int_k 
\overline\Gamma_{\!0\,\alpha}(k) S(k_1) 
\overline\Gamma_{\!0}^\mu(q)
S(k_2)\overline\Gamma_{\!0}^\alpha(k) V(q^2)V^2(k^2)\Delta(k^2)\,,\nonumber\\
\nonumber\\
\displaystyle c_2^\mu & \displaystyle = & \displaystyle ic_b\int_k \overline\Gamma_{\!0}^\alpha(\ell) S(k_2)\overline\Gamma_{\!0}^\beta(k){\overline\g}^{\mu\alpha\beta}(q,\ell,-k)\Delta(\ell^2)\Delta(k^2)V(\ell^2)V(k^2)\,,
\eea
where  
\be\label{eq:int_measure}
\int_{k} := \frac{1}{(2\pi)^4} \int \!\!{\rm d}^4 k \,
\ee
denotes the suitably regularized integration over virtual momenta. In addition, we have set $k_1=p+k$, $k_2=r+k$ and $\ell=k-q$, and  
defined \mbox{$c_a =-g^2(C_f-C_A/2)$}, $c_b=g^2C_A/2$, 
where $C_f$ and $C_A$
are the Casimir eigenvalue
of the fundamental and adjoint representations, respectively;
for $SU(3)$, $C_f=4/3$ and 
$C_A =3$.
The quantity 
$\overline{\g}^{\, \mu \alpha \beta}$ appearing in the 
integrand of $(c_2)$
denotes the 
transversely-projected three-gluon vertex
($r\to\-k$, $p \to \ell$), 
\be
\overline{\g}^{\, \mu \alpha \beta}(q,r,p) =  P_{\mu'}^\mu(q)  P_{\alpha'}^{\alpha}(r)  P_{\beta'}^\beta(p) 
\g^{\,\mu'\! \alpha' \!\beta'}(q,r,p) \,,
\ee
whose color factor $f^{abc}$  has been 
absorbed into the  $c_b$. 


Owing to the planar degeneracy property of this vertex \cite{Blum:2014gna,Eichmann:2014xya,Ferreira:2023fva,Aguilar:2023qqd,Pinto-Gomez:2022brg,Pinto-Gomez:2024mrk}, the tree-level structure alone
\be\label{eq:planardeg}
\g^{\mu\alpha\beta}(q,r,p)=L_{sg}(s^2)\left[g^{\alpha\beta}(r-p)^\mu+g^{\mu\beta}(p-q)^\alpha+g^{\mu\alpha}(q-r)^\beta\right]\,,
\ee
with \mbox{$\displaystyle s^2=\frac{1}{2}(q^2+r^2+p^2)$}, provides an excellent description of the full three-gluon vertex. The functional form of the form factor $L_{sg}(s^2)$ \cite{Athenodorou:2016oyh,Duarte:2016ieu,Boucaud:2017obn, Vujinovic:2018nqc,Pinto-Gomez:2022brg,Aguilar:2019uob,Aguilar:2021lke,Aguilar:2021okw,Pinto-Gomez:2024mrk} used in the present analysis is discussed in \sect{sec:num}. 

Note finally that, 
as was shown in \cite{Aguilar:2024ciu}, 
the contribution of the Abelian 
graph to $\lambda_i$ 
is practically negligible,
accounting for only 2 per mil of the full answer. Therefore, we will simplify  the ensuing analysis by omitting this 
graph entirely.  

\section{The soft-gluon limit}\label{sec:sglim}

For the study of the quark-gluon vertex 
in the complex plane we will 
restrict ourselves to 
a typical kinematic configuration known in the literature as the 
``{\it soft-gluon}'' limit, defined by 
$q = 0$,  and $r=p$; the only remaining momentum will be subsequently complexified,
by setting $p^2 \to z \in  \mathbb{C}$.
This particular kinematic choice permits one to distribute $p$ inside the diagrams 
$(c_1)$ and $(c_2)$ of \fig{fig:greenqgsde} such that the gluon propagators  
carry only the Euclidean integration momentum $k$. 

The  quark-gluon vertex 
considered in this work has the general form
\be
\label{Gsg}
\overline{\Gamma}^{sg}_{\!\mu}(q,r,-p)=\sum_{i=1}^8\lambda_i^{sg}(p^2)\bar{\tau}_i^\mu(r,-p)\,,
\ee
with the form factors defined as 
\begin{align}\label{eq:ffsdef}
    \lambda_i^{sg}(p^2)&=\lim_{q\to 0}\lambda_i(q,r,-p)
    \,,\qquad & \lambda_i(q,r,-p)&=\textrm{Tr}\left[\mathcal{P}_{\!i}^\mu(q,r,-p)\overline{\Gamma}_{\!\mu}(q,r,-p)\right]\,,
\end{align}
where the closed form of the projectors $\mathcal{P}_{\!i}(q,r,-p)$ is given in Eq.\,(3.9) of \cite{Aguilar:2024ciu}. 
The form of \1eq{Gsg}
is to be contrasted with the 
case where the limit $q \to 0$ is taken also 
at the level of the basis elements 
$\tau_{i}^\nu(r,-p)$, thus reducing their number down to three, 
see, \eg Eq.~(A8)~in~\cite{Aguilar:2024ciu}. 

Turning to \1eq{eq:vertexsde}, and 
retaining only the dominant 
contribution $(c_2)$, one has 
\be\label{eq:lambdasgdef}
\lambda_i^{sg}(p^2)=\delta_{i1}+
\lambda_{i,\s{Q}}^{sg}(p^2) \,,  \qquad 
\lambda_{i,\s{Q}}^{sg}(p^2) :=
\lim_{q\to 0}\textrm{Tr}\left[\mathcal{P}_{\!i}^\mu(q,r,-p)P_{\mu\nu}(q)c_2^\nu(q,r,-p)\right]\,.
\ee
Passing to Euclidean space following standard rules, see \eg Sec.~IV.~B and App.~A in \cite{Aguilar:2024ciu}, one has
\be\label{eq:liSDE1}
\lambda_i^{sg}(p^2)=\delta_{i1}+c_b\lim_{q\to 0}\int_E \textrm{Tr}\left[\mathcal{P}_{\!i}^\mu(q,r,-p)\overline\Gamma_{\!0}^\alpha(\ell) S(k_2)\overline\Gamma_{\!0}^\beta(k){\overline\g}^{\mu\alpha\beta}(q,\ell,-k)\Delta(\ell^2)\Delta(k^2)V(\ell^2)V(k^2)\right]\,.
\ee
with $k_2=r+k$, and
\be
\int_E:=\frac{1}{2(2\pi)^3}\int_{0}^\infty \!\!k^2 dk^2\int_0^\pi\!\!\sin^2\!\omega \,d\omega \int_0^\pi\!\!\sin\phi \,d\phi\,,
\ee

Employing a typical parametrization 
for the momenta (see, \eg Eqs.~(3.18)~and~(3.19) of \cite{Aguilar:2016lbe}), the relevant inner products become 
\be
\begin{array}{rclcrcl}
    \displaystyle p\cdot r & \displaystyle = & \displaystyle \sqrt{p^2r^2}\cos\theta_{rp}\,,&\qquad\qquad&\displaystyle p\cdot k & \displaystyle = & \displaystyle \sqrt{p^2k^2}\cos\omega\,,\\
    \\
    \multicolumn{7}{c}{r\cdot k=\sqrt{r^2k^2}\left(\cos\theta_{rp}\cos\omega+\sin\theta_{rp}\sin\omega\cos\phi\right)}\,.
\end{array}
\ee
Then, the soft-gluon limit is implemented by first taking $r\to p$, followed by $\theta_{rp}\to 0$. This limiting procedure requires a careful Taylor expansion, 
due to the presence of divergent 
terms in the expression 
in square brackets on the r.h.s. of \1eq{eq:liSDE1}; thus, one 
finally obtains well-defined 
expressions for all form factors $\lambda^{sg}_i(p^2)$.

 In particular, 
we obtain the integral expressions  
\bea\label{eq:lmbds}
    \lambda^{sg}_i(p^2)  & = &  \delta_{i1}+c_b\!\int_{k,\omega} \mathcal{R}(k^2)K_{i} a(k_1^2)\,,\qquad\textrm{for }i=1,6,7\,;\nonumber\\
    \nonumber\\
    \lambda^{sg}_5(p^2) &= &c_b\int_{k,\omega} \mathcal{R}(k^2)K^a_5 a(k_1^2)+c_b\int_{k,\omega} K^b_5 \left[\frac{1}{2}\mathcal{R}'(k^2)a(k_1^2)+\mathcal{R}(k^2)a'(k_1^2)\right]\,;\nonumber\\
    \nonumber\\
    \lambda^{sg}_i(p^2) &= & c_b\!\int_{k,\omega} \mathcal{R}(k^2)K_{i} b(k_1^2)\,,\qquad\qquad\textrm{for }i=2,3,5,8\,;
\eea
where we use $f'(x) := df(x)/dx$, and  
$K_i: =K_i(k^2,p^2,\omega)$.   
The two-dimensional integration is defined as  
\be
\int_{k,\omega}:=\frac{1}{2(2\pi)^3}\int_0^\infty k^2 dk^2\int_0^\pi \sin^2\!\omega \, d\omega\,, \label{int_ang}
\ee
and we have introduced the key dynamical quantity 
\be\label{eq:calRdef}
{\cal R}(k^2) =V^2(k^2)\Delta^2(k^2)L_{sg}(k^2)\,,
\ee
together with 
\be
a(p^2) := \frac{A(p^2)}{p^2 A^2(p^2) + B^2(p^2)} \,, \qquad 
b(p^2) := \frac{B(p^2)}{p^2 A^2(p^2) + B^2(p^2)} \,.
\ee
Finally, the kernels are given by
\be
\begin{aligned}[b]
    &\displaystyle K_1 = \displaystyle -\frac{4}{3}(3k^2+2\sqrt{k^2p^2}\cos\omega)\sin^2\!\omega\,,\quad K_2 = -6\sqrt{\frac{k^2}{p^2}}\cos\omega\,, \quad K_3 =0\,,\\[5pt]
    &\displaystyle K_4 = \displaystyle 4(1+\cos^2\!\omega)\,,\qquad  K^a_5 = -\frac{4}{3}\frac{k^2}{p^2}\cos^2\!\omega(3\sqrt{k^2p^2}+2\sqrt{p^2}\cos\omega)\sin^2\!\omega\,,\\[5pt]
    &\displaystyle K_5^b = \displaystyle -\frac{1}{3p^2}(3k^2+2p^2(3+2\cos^4\!\omega)+10\sqrt{k^2p^2}(5-\cos^2\!\omega)\cos\omega -4(3k^2-2p^2)\cos^2\!\omega)\,,\\[5pt]
    &\displaystyle K_6 = \displaystyle \frac{1}{3p^2}(-3k^2+\sqrt{k^2p^2}\cos\omega+12k^2\cos^2\!\omega+8\sqrt{k^2p^2}\cos^3\!\omega)\,,\\[5pt]
    &\displaystyle K_7  = \displaystyle -8\sqrt{\frac{k^2}{p^2}}\cos\omega-4(1+\cos^2\!\omega)\,,\qquad K_8 = -\frac{2}{3p^2}(4\cos^2\!\omega-1)\,.
\label{kernels}
\end{aligned}
\ee

Note that the integral for $\lambda_5^{sg}$ involves derivatives of the scalar functions, as a result of the aforementioned 
Taylor expansion. 
In addition, we observe that the kernel $K_3$, and hence  $\lambda_3^{sg}$, 
vanish identically, in compliance 
with charge conjugation symmetry \cite{Aguilar:2024ciu}. 
Finally, we remark that  
when the chirally symmetry is unbroken, $B(p^2)=0 \implies  b(p^2)=0$, the 
chiral symmetry breaking form factors $\lambda_{2,3,4,8}$ vanish identically,
as expected. 

\section{Analytic structure of the vertex integrals}\label{sec:anstruct}

Before embarking on the numerical analysis of the $\lambda^{sg}_i(p^2)$ in the complex plane, we review the systematics of the Wick rotation in loop integrals, and the domain where the standard Euclidean space integral of \1eq{eq:lmbds} is valid. Moreover, we consider a perturbative toy model for the calculation of vertex-type integrals, which will serve as a benchmark for the numerical methods employed in the nonperturbative case.

\subsection{General considerations}

The computation of the form factors $\lambda^{sg}_i(p^2)$ in 
Minkowski space amounts to the evaluation of integrals of the form
\be\label{eq:typeInt}
I(p^2)=\int_k f(k_0, \kvec, p) = \frac{1}{(2\pi)^4}\int \! d^3\kvec \int_{-\infty}^\infty \!\!\!\! d k_0 \, f(k_0,\kvec,p) \,,
\ee
where $f(k_0,\kvec,p)$ is a scalar function.
Note that we explicitly decompose the integration momentum \mbox{$k = (k_0,\kvec)$} into temporal and spatial parts (boldface letters). Moreover, since in this section we use both Minkowski and Euclidean momenta, the latter will be distinguished by an index ``$\rm{E}$". For a purely Euclidean perspective, see~\cite{Huber:2022nzs,Huber:2023uzd}.

The main difficulty in evaluating \1eq{eq:typeInt} arises from the existence of singularities in the integrand, $f(k_0, \kvec, p)$. In particular, the fits and models used as external inputs in \sect{sec:num} for $\Delta(k^2)$, $\Ls(k^2)$, $V(k^2)$, and $S(k)$, contain complex-conjugate poles. Since the relation between the Minkowski and Euclidean integrals depends on the analytic structure of the integrand $f(k_0, \kvec, p)$ in the complex $k_0$ plane, the presence of such non-analyticities limits the domain of complex values of $p^2$ for which the form factors 
$\lambda^{sg}_i(p^2)$
can be evaluated directly through \1eq{eq:lmbds}.

\begin{figure}[t]
\centering
\includegraphics[width=0.3\textwidth]{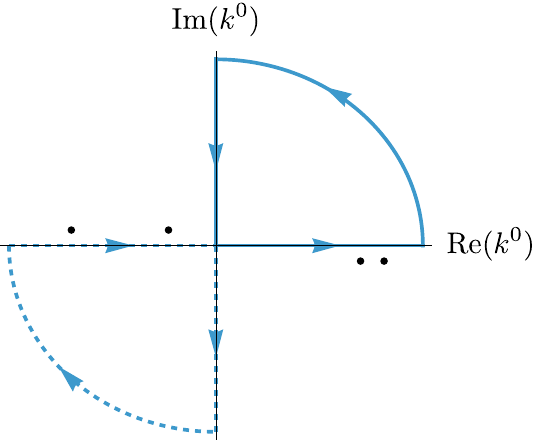}\hfil\includegraphics[width=0.3\textwidth]{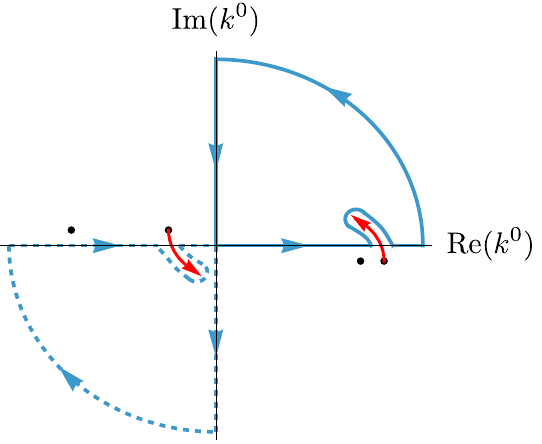}\hfil\includegraphics[width=0.3\textwidth]{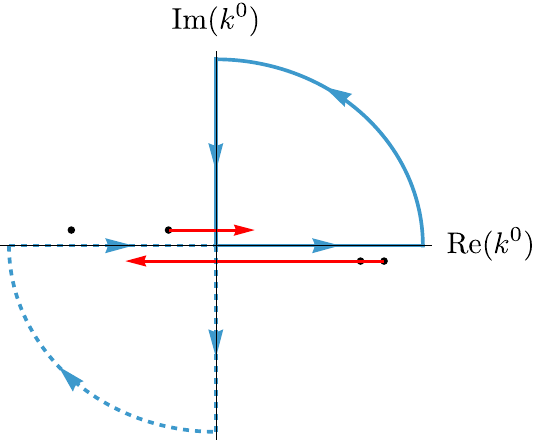}
\caption{Left: The function $f(k_0, \kvec, p)$ is analytic on and inside the contours (blue solid and blue dashed lines), and 
\1eq{int_euc} is valid. Center: The poles move vertically into the first and/or third quadrants (red arrows indicate the motion of the poles). The contours are deformed to keep $I(p^2)$ well defined, and \1eq{int_euc} is still valid. Right: Poles move horizontally into the first and/or third quadrants; their residues must be taken into account, and \1eq{int_euc} is modified to \1eq{euc_int}. }
\label{fig:Contour_All}
\end{figure}

To begin with, suppose that for some domain of values of $p$ the $f(k_0, \kvec, p)$ is analytic in the first and third quadrants of the complex $k_0$ plane, while containing poles in the other two quadrants, as shown on the left panel of \fig{fig:Contour_All}. Then, Cauchy's theorem implies
\be 
\oint_C f(k_0, \kvec, p) = \int_0^{\infty} \!\!\!\!dk_0  f(k_0, \kvec, p) - i \!\int_0^\infty \!\!\!\!d k_0^\srm{E} f(ik_0^\srm{E},\mathbf{k},p) = 0 \,,
\ee
where $C$ is the contour shown as a blue solid line encompassing the first quadrant on the left panel of \fig{fig:Contour_All}, and the second integral is over the imaginary $k_0 = i k_0^\srm{E}$ axis. Evidently, the above result requires that $f(k_0, \kvec, p)\to 0$ sufficiently fast at $|k_0|\to \infty$ for the integral over the quarter circle at infinity to vanish, as is usually the case. Operating similarly in the third quadrant, with the contour shown as a blue dashed line on the left panel of \fig{fig:Contour_All}, and restoring the $d^3\kvec$ integration, one obtains
\be 
I(p^2)=\int_k f(k_0, \kvec, p) = i\!\int\!\! \frac{d^4k_\srm{E}}{(2\pi)^4} f(ik_0^\srm{E},\kvec^\srm{E},p) \,, \label{int_euc}
\ee
where $\kvec = \kvec^\srm{E}$. Thus, in this case, $I(p^2)$ can be evaluated through direct integration in Euclidean space, \ie through the standard Wick rotation $k_0\to i k_0^\srm{E}$. Note that in the above expression, $p$ is still a Minkowski space momentum. 

As $p$ is varied, however, the poles of $f(k_0, \kvec, p)$ may move into the first and/or third quadrants of the complex $k_0$ plane. Depending on whether the poles move into these quadrants vertically, \ie crossing the real $k_0$ axis, or horizontally, the equality of the Minkowski and Euclidean integrals will be invalidated.

When a pole moves vertically into a different quadrant as $p$ is varied, the real $k_0$ integration becomes ill-defined when the pole is on top of the axis. In this case, $I(p^2)$ must be \emph{defined} by analytic continuation. This can be achieved by continuously deforming the integration contour to prevent the pole from crossing it, as shown on the central panel of \fig{fig:Contour_All}. With this deformation, $f(k_0,\kvec,p)$ remains analytic within and on the closed contours shown on the central panel of \fig{fig:Contour_All}. Hence, Cauchy's theorem guarantees again the equality of the Euclidean integral and the analytically continued Minkowski one, \ie \1eq{int_euc}.

On the other hand, when poles move horizontally and cross the imaginary $k_0$ line, as shown on the right panel of \fig{fig:Contour_All}, the Minkowski integral is not touched by the pole, and $I(p^2)$ remains analytic without any contour deformation. However, since the poles now appear in either the first or third quadrants, the relation between the Euclidean and Minkowski integrals is modified.

Specifically, applying the residue theorem to the contour in the first quadrant on the right plot of \fig{fig:Contour_All}, we get
\be 
\int_0^{\infty} \!\!\!\!dk_0 f(k_0,\kvec,p) - i \!\int_0^\infty \!\!\!\!d k_0^\srm{E} f(ik_0^\srm{E},\kvec,p) = 2\pi i \!\!\!\!\sum_{\substack{\text{poles in}\\\text{1}^\text{st}\text{ quad.}}}\!\!\!\!\text{Res} f(k_0,\mathbf{k},p)\,.
\ee
Similarly, the third quadrant yields
\be 
\int_{-\infty}^0 \!\!\!\!dk_0 f(k_0,\mathbf{k},p) - i \!\int_{-\infty}^0 \!\!\!\!d k_0^\srm{E} f(ik_0^\srm{E},\mathbf{k},p) = - 2\pi i \!\!\!\!\sum_{\substack{\text{poles in}\\\text{3}^\text{rd}\text{ quad.}}}\!\!\!\!\text{Res} f(k_0,\mathbf{k},p) \,,
\ee
where the minus sign in the residue is due to the clockwise orientation of the contour. Hence, restoring the $d^3\mathbf{k}$ integral, and the factors of $2\pi$,
\be 
I(p^2) = i \int \!\! \frac{d^4k_\srm{E}}{(2\pi)^4}  f(ik_0^\srm{E},\mathbf{k}_\srm{E}, p) + i\!\int \!\!\frac{d^3\mathbf{k}}{(2\pi)^3} \left[ \sum_{\substack{\text{poles in}\\\text{1}^\text{st}\text{ quad.}}}\!\!\!\!\text{Res} f(k_0,\mathbf{k},p) - \!\!\!\! \sum_{\substack{\text{poles in}\\\text{3}^\text{rd}\text{ quad.}}}\!\!\!\!\text{Res} f(k_0,\mathbf{k},p) \right] \,. \label{euc_int}
\ee
Therefore, in this case, the Euclidean integral no longer yields the complete $I(p^2)$, which must be computed taking into account the residues of the poles in the first and third quadrants.

Moreover, since the poles will generally occur at complex values of $k_0$, the determination of the residues requires knowledge of the integrand, and hence the nontrivial dressing functions appearing in it, for complex arguments. Because the structure of these dressing functions in the complex plane is still poorly known, direct calculation of the $\lambda^{sg}_i(p^2)$ will be limited, in practice, to the domain where the Euclidean and Minkowski integrals coincide.

As a final remark, we note that when the poles move horizontally into the first/third quadrants, one could also deform the imaginary $k_0$ integration to keep the poles out. That, however, is neither necessary, since $I(p^2)$ defined from the Minkowski integral is already analytic under such movement, nor would it change the conclusion in \1eq{euc_int}. Indeed, a deformation of the integration contour over the imaginary $k_0$ would amount to modifying the Euclidean integral, such that its standard form would no longer be valid. Moreover, it is straightforward to show that the integral over the deformed contour would yield precisely the standard Euclidean integral plus the residue terms in \1eq{euc_int}.

\subsection{A toy model}

As an illustration of the above discussion, we consider a vertex-type of integral 
(triangle graph) 
that can be computed exactly. Since 
a variety of functional studies~\cite{Alkofer:2003jj,Fischer:2008sp,Dorkin:2013rsa,Windisch:2016iud,Eichmann:2016yit, Miramontes:2019mco,Gao:2024gdj,Alkofer:2026vux} of the quark propagator encounter a pair of complex-conjugate poles, we will focus on the case where the integrand contains such structures. 

Specifically, consider 
\begin{align}
f(k_0,\kvec,p) =\frac{i}{(k^2 - m^2 + i \epsilon)^2}\left[ \frac{1}{(k-p)^2 - M^2 + i \epsilon} +  \frac{1}{(k-p)^2 - M^*{}^2 + i \epsilon} \right]  \,,
\end{align}
consisting of two tree-level propagators with mass $m$, assumed for simplicity to be real, and a propagator containing a pair of complex conjugate masses, $M$ and $M^\star$. In addition, we assume that $\textrm{Re}(M^2)> 0$, such that the corresponding propagator is analytic in the negative (positive) half-plane of the complexified $p^2$ in Minkowski (Euclidean) space. Therefore, since the masses appear squared, we can write without loss of generality $M = |M|e^{i \varphi_{\scalebox{0.4}{$M$} }}$, with $\varphi_{\!\scriptscriptstyle M} \in [0, \pi/4]$. Finally, the factor of $i$ is introduced to make the resulting $I(p^2)$ real and positive for space-like $p$.

To carry out the computations, it is convenient to split the integral into two pieces, each containing one of the complex-conjugate poles, \ie
\be
I(p^2) = J(p^2,M^2) + J(p^2,M^*{}^2) \,, \label{I_from_J}
\ee
where
\be 
J(p^2,M^2) := \int_k \bar{f}(k_0,k,p) \,, \quad \bar{f}(k_0,k,p) := \frac{i}{(k^2 - m^2 + i \epsilon)^2[(k-p)^2 - M^2 + i \epsilon]} \,.
\ee

By Lorentz invariance, we can analyze the system in the rest-frame of the particle with nonzero momentum $p$, such that $p = (p_0, \mathbf{0})$. Then, $\bar{f}(k_0,\kvec,p)$ reads
\begin{align} 
\bar{f}(k_0,\kvec,p)  =&\, \frac{i}{(k_0^2 - \mathbf{k}^2 - m^2 + i\epsilon )^2(k_0^2 - \mathbf{k}^2 - 2 k_0 p_0 + p_0^2 - M^2 + i\epsilon)} \,. \label{f0bar}
\end{align}
In addition, since $I(p^2)$ depends only on $p^2$ and displays the Schwarz reflection property, $I(z^\star) = I^\star(z)$, we can restrict our analysis to $\textrm{Im}(p_0)\geq0$, \ie $p_0 = |p_0| e^{i\varphi}$, with  $\varphi \in [0,\pi/2]$.

With the above parametrization, $\bar{f}(k_0,\kvec,p)$ has poles at
\be 
k_{0,1}^{\pm} = \pm \sqrt{\mathbf{k}^2 + m^2} \mp i \epsilon \,, \label{pole1}
\ee
originating in the propagators with mass $m$, and 
\be
k_{0,2}^{\pm} = p_0 \pm \sqrt{\mathbf{k}^2 + M^2 } \mp i \epsilon  \,,  \label{pole2}
\ee
in the propagator with complex mass $M$. 

The poles at $k_{0,1}^{\pm}$ are in the second and fourth quadrants of the $k_0$ axis, respectively, independently of $p$. Thus, they never appear in the sums on the right-hand side of \1eq{euc_int}.  On the contrary, the poles $k_{0,2}^{\pm}$ may move from one quadrant to another, depending on the values of $p$, $M$, and $\kvec$.

\begin{figure}[t]
\centering
\includegraphics[scale=1]{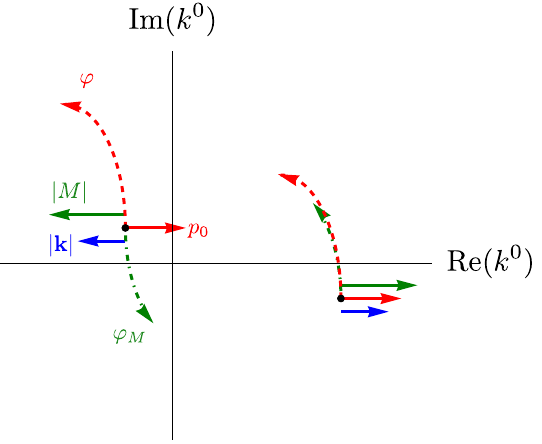}
\caption{The positions of the poles $k_{0,2}^{-}$ (left) and $k_{0,2}^{+}$ (right) of \1eq{pole2} for real $M > p_0>0$ are illustrated as black dots. They are displaced from the $k_0$ axis by the $i\epsilon$ prescription, exaggerated in the figure for visibility. The arrows indicate the direction of their motions as we increase any one of the variables $|\kvec|$ (blue solid), $|M|$ (green solid), $\varphi_{\!\scriptscriptstyle M}$ (green dashed), $|p_0|$ (red solid), and $\varphi$ (red dashed).}
\label{fig:Pole_motion}
\end{figure}

To analyze the latter case, note that for real $M$ ($\varphi_{\!\scriptscriptstyle M} = 0$) and a time-like $p$ with \mbox{$0 < p_0 < M~(\varphi = 0)$}, the pole $k_{0,2}^{-}$ is in the second quadrant, and $k_{0,2}^{+}$ in the fourth, as shown in \fig{fig:Pole_motion}; in this case, the Euclidean and Minkowski integrals coincide.

As $p_0$ is rotated towards the imaginary axis, \ie for $\varphi > 0$, the pole $k_{0,2}^{-}$ stays in the second quadrant, while $k_{0,2}^{+}$ crosses the $k_0$ axis moving into the first, as shown by the red-dashed arrows in \fig{fig:Pole_motion}. In this case, the contour has to be deformed to maintain $I(p^2)$ well-defined, as on the central panel of \fig{fig:Contour_All}, and the Euclidean and Minkowski integrals remain equal.

On the other hand, as $p_0$ is increased, both poles move towards the right (red solid arrows in \fig{fig:Pole_motion}). In particular, for $p_0 > M$, the pole $k_{0,2}^{-}$ may move into the first quadrant, depending on $\kvec^2$. When this happens, the Euclidean and Minkowski integrals are no longer equal, being related by \1eq{euc_int} instead of \1eq{int_euc}. 

Similarly, we can analyze the dependence of $k_{0,2}^{\pm}$ on the mass $M$. As shown with the green solid arrows in \fig{fig:Pole_motion}, increasing $|M|$ moves both poles away from the imaginary axis, increasing the domain of validity of the Euclidean integral. When the mass turns complex ($0 < \varphi_{\!\scriptscriptstyle M} < \pi/4$, green dot-dashed lines), both poles cross the $k_0$ axis, entailing a contour deformation. In addition, both $k_{0,2}^{\pm}$ move slightly closer to crossing horizontally to a different quadrant. Thus, the Euclidean integral possesses a more 
restricted domain of validity with a complex mass than with a real mass of equal modulus $|M|$.

In general, the poles move horizontally into a different quadrant if and only if
\be 
\textrm{Re}(k_{0,2}^{+}) < 0 \,, \qquad \textrm{Re}(k_{0,2}^{-}) > 0 \,,
\ee
which implies
\be 
\textrm{Re}^2(p_0) > \textrm{Re}^2(\sqrt{\kvec^2 + M^2}) \,. \label{domain0}
\ee
Then, since $\kvec^2>0$ displaces the poles that are on the left further to the left, and the poles on the right further to the right (blue arrows in \fig{fig:Pole_motion}), the $k_{0,2}^{\pm}$ can only enter horizontally into a different quadrant if they do so for $\kvec = 0$. Hence, for
\be 
\textrm{Re}^2(p_0) < \textrm{Re}^2( M ) = |M|^2\cos^2 \!\!\varphi_{\scriptscriptstyle M} \,, \label{domain1}
\ee
the poles cannot cross horizontally to a different quadrant for any $\kvec$;
thus, the above inequality defines the domain of validity of
\1eq{int_euc}. 

\begin{figure}[t]
\centering
\includegraphics[scale=1]{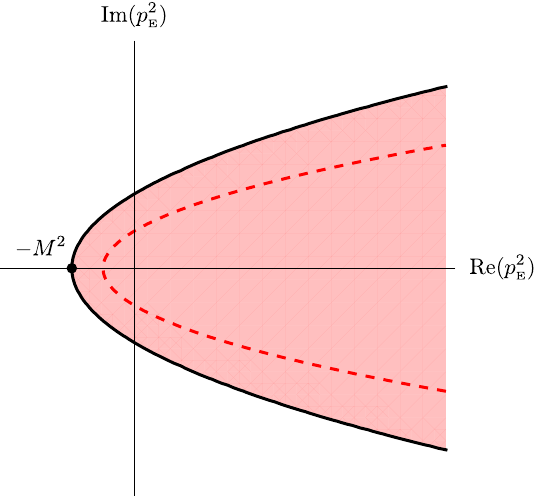}
\caption{Domain of validity of the standard Euclidean integral in the complex $p^2_\srm{E}$-plane for fixed $|M|$. The shaded area, delimited by a black line corresponds to the case of $M\in\mathbb{R}$; the red-dashed contour delimits the $p^2_\srm{E}$ region accessible for a complex mass.}
\label{fig:q2reg}
\end{figure}

Since practical calculations are generally performed in Euclidean space, and $I(p^2)$ depends only on $p^2$, it is convenient to express \1eq{domain1} in terms of the Euclidean squared momentum, $p_\srm{E}^2 = - p^2$, such that $p_0 = \sqrt{ - p_\srm{E}^2 }$. Then, writing $p_\srm{E}^2 = x + i y$, we have
\be 
p_0 = \sqrt{ - x - i y }  = \sqrt{\frac{\sqrt{ x^2 + y^2 } - x}{2} } + i \, \textrm{sgn}(y)\sqrt{\frac{\sqrt{ x^2 + y^2 } + x}{2} } \,.
\ee
With a little algebra, \1eq{domain1} amounts to
\be\label{eq:parab}
y^2 < 4 |M|^2\cos^2 \!\!\varphi_{\!\scriptscriptstyle M}  \left( x + |M|^2\cos^2\!\! \varphi_{\!\scriptscriptstyle M}  \right) \,,
\ee
which defines a parabolic region in the complex $p_\srm{E}^2$-plane (see \fig{fig:q2reg}) delimited by the parabola \mbox{$y^2 = 4 |M|^2\cos^2 \varphi_{\!\scriptscriptstyle M}\left( x + |M|^2\cos^2\!\! \varphi_{\!\scriptscriptstyle M}  \right)$}, with apex at $p_\srm{E}^2 = - |M|^2\cos^2 \!\!\varphi_{\!\scriptscriptstyle M} $ and focus at the origin. Note that in the $M = 0$ limit, the domain shrinks to the space-like half-line, $p_\srm{E}^2 > 0$.

For fixed nonzero $|M|$, recalling that  $\varphi_{\!\scriptscriptstyle M} \in [0, \pi/4]$, the domain of validity of the Euclidean integral is largest when $M$ is real ($\varphi_{\!\scriptscriptstyle M}= 0$), and smallest for $\varphi_{\!\scriptscriptstyle M} = \pi/4$, in which case $M^2$ is purely imaginary. The boundaries of the largest and smallest domains are represented in \fig{fig:q2reg} by a black-continuous and a 
red-dashed line, respectively. 

\subsection{Exact evaluation}\label{subsec:exact}

For the model discussed above, the integral $I(p^2)$ of \1eq{eq:typeInt} can be computed exactly. Indeed, for real $p^2$ and $M^2$, the integral $J(p^2,M^2)$ can be evaluated explicitly using Feynman parametrization. The result can then be analytically continued to complex momenta and masses, and combined through  \1eq{I_from_J} to obtain $I(p^2)$. Expressed in terms of the Euclidean squared momentum, $p_\srm{E}^2 = - p^2$, the final expression reads
\be\label{eq:Ires}
I(-p^2_\srm{E})= \frac{1}{16\pi^2p^2_\srm{E}}\left[ 2 \ln\left(\frac{|M|}{m}\right) + \Lambda(p_\srm{E}^2,M,m)+\Lambda(p_\srm{E}^2,M^\ast,m) \right]\,,
\ee
where
\be
\Lambda(p_\srm{E}^2,M,m)=\frac{p^2_\srm{E} + m^2-M^2}{\sqrt{\lambda(-p^2_\srm{E},M^2,m^2)}}\ln\left(\frac{p^2_\srm{E} + m^2+M^2+\sqrt{\lambda(-p^2_\srm{E},M^2,m^2)}}{2mM}\right)\,,
\ee
and ${\displaystyle \lambda(x,y,z):= x^{2}+y^{2}+z^{2}-2xy-2yz-2zx}$ is the K\"allen function. 

Note that the minus sign in the argument of $I(-p_\srm{E}^2)$ in \1eq{eq:Ires} traces back to its original definition in Minkowski space in \1eq{eq:typeInt}.  In nonperturbative studies carried out entirely in Euclidean space, it is usual to define the function under study with argument $p_\srm{E}^2$, rather than $-p_\srm{E}^2$, which amounts to a redefinition of the function, \eg $\widetilde{I}(p_\srm{E}^2) := I(-p_\srm{E}^2)$. Exceptionally, in this section, we retain all of the original signs and indices ``E'', since both Minkowski and Euclidean spaces are referred to.

Given that the implementation of the 
Feynman parametrization is not possible 
in a  nonperturbative context, it is 
instructive to evaluate $I(p^2)$ 
using the method discussed in the previous 
subsection, namely use of \1eq{int_euc}
and, when required, the 
corresponding residue term. This computation 
allows us to benchmark methods that 
may be applied nonperturbatively  
against an exact result, namely \1eq{eq:Ires}.  

Within the parabolic region of \1eq{eq:parab}, the split integral $J(p^2)$ can be evaluated through the standard Euclidean expression, \ie \1eq{int_euc}, which reads explicitly
\be 
J(p^2, M^2)= J_\srm{E}(-p_\srm{E}^2, M^2) := \int\!\! \frac{d^4k_\srm{E}}{(2\pi)^4} \frac{1}{(k_\srm{E}^2 + m^2)^2( k_\srm{E}^2 + 2 i k^\srm{E}_0 \sqrt{- p_\srm{E}^2} + p_\srm{E}^2 + M^2 )} \,. \label{int_euc_exp}
\ee
Then, using hyperspherical coordinates, we can recast the above equation in terms of the integral measure defined in \1eq{int_ang},
\be 
J_\srm{E}(-p_\srm{E}^2, M^2) = 2\int_{k_\srm{E},\omega} \frac{1}{(k_\srm{E}^2 + m^2)^2(k_\srm{E}^2 + 2 i k^\srm{E}_0 \sqrt{- p_\srm{E}^2} \cos \omega + p_\srm{E}^2 + M^2 )} \,. \label{int_euc_ang}
\ee
In particular, for space-like momenta, $p_\srm{E}^2>0$, we have that $\sqrt{- p_\srm{E}^2} = i|p_\srm{E}|$, and the above expression attains the more recognizable form
\be 
J_{\srm E}(-p_\srm{E}^2, M^2) = 2 \int_{k_\srm{E},\omega} \frac{1}{(k_\srm{E}^2 + m^2)^2(k_\srm{E}^2 - 2  |k_\srm{E}| |p_\srm{E}| \cos \omega + p_\srm{E}^2 + M^2 )} \,.
\ee

Outside the parabolic region in \1eq{eq:parab}, the poles in the first and third quadrants must be taken into account, as in \1eq{euc_int}; for the present simple model, this can be done explicitly.

Since we consider $\textrm{Im}(p_0)>0$, the only pole that needs to be accounted for is $k_{0,2}^{-}$, when it moves into the first quadrant. However, even for $p_\srm{E}^2$ outside the parabola, the pole only moves into the first quadrant when \1eq{domain0} is satisfied. With some algebra, that inequality can be recast as
\be 
\kvec^2 < r_0^2 \,, \label{k_less}
\ee
where 
\be 
r_0^2 := \textrm{Re}^2\left( \sqrt{-p_\srm{E}^2} \right) - \textrm{Re}^2(M^2) = \textrm{Re}^2\left( \sqrt{-p_\srm{E}^2} \right) - |M|^2\cos(2\varphi_{\scriptscriptstyle M}) \,, \label{radius}
\ee
which implies that the pole only moves horizontally into the first quadrant for $\kvec$ inside a sphere of radius $r_0$.

Then, the residue of $k_{0,2}^{-}$ can be computed explicitly from \1eq{f0bar}, yielding
\be 
\sum_{\substack{\text{poles in}\\\text{1}^\text{st}\text{ quad.}}}\!\!\!\!\text{Res} f(k_0,\kvec,p) =  \frac{-i\Theta(r_0^2  - \kvec^2)}{2(M^2 - m^2 - 2p_0 \sqrt{r^2 + M^2} + p_ 0^2 )^2\sqrt{ r^2 + M^2} } \,, \label{res_term}
\ee
where the Heaviside theta function, $\Theta(x)$, enforces the condition in \1eq{k_less}. 

At this point, the $d^3\kvec$ integral can be evaluated using spherical coordinates. Noting that \1eq{res_term} only depends on $\kvec^2$, we have, with $r:=|\kvec|$,
\be 
J_\srm{R}(-p_\srm{E}^2, M^2) = \frac{1}{(2\pi)^2}\!\int_0^{r_0} \!\! dr \frac{r^2}{\Big[M^2 - m^2 - 2 \sqrt{- p_\srm{E}^2(r^2 + M^2)} - p_\srm{E}^2 \Big]^2\sqrt{ r^2 + M^2} } \,. \label{res_term_fin}
\ee
Note that if the upper limit of integration were infinity, $J_\srm{R}$ would diverge logarithmically. Instead, the finite radius of integration guarantees the finiteness of $J_\srm{R}$, manifesting the known fact that ultraviolet divergences are purely Euclidean~\cite{Collins:1984xc}.

Note also that the integrand of \1eq{res_term_fin} has a pole at $r = 0$ ($\kvec = 0$) when $p^2_\srm{E} = - (M + m)^2$. At this squared momentum, the poles $k_{0,2}^-$ and $k_{0,1}^+$ pinch the $k_0$ contour, which cannot be deformed in any way to avoid the singularity.  As a result, $J_\srm{R}(-p_\srm{E}^2, M^2)$, and hence $I(-p^2_\srm{E})$, has a branch point, corresponding to the production threshold~\cite{Huber:2022nzs,Huber:2023uzd}.

Finally, we combine the above results to obtain $I(-p_\srm{E}^2)$ through \1eq{I_from_J}. Denoting by $I_\srm{E}(-p_\srm{E}^2)$ the contribution from the standard Euclidean integrals and $I_\srm{R}(-p_\srm{E}^2)$ the residue terms, \ie
\be 
I_\srm{E}(-p_\srm{E}^2) = J_\srm{E}(-p_\srm{E}^2, M^2) + J_\srm{E}(-p_\srm{E}^2, M^*{}^2) \,,
\ee
and similarly for $I_\srm{R}$, the final result for $I(-p^2_\srm{E})$ is given by
\be 
I(-p^2_\srm{E})=\begin{cases}
			I_\srm{E}(-p_\srm{E}^2)\,, & \text{if $y^2 < 4 |M|^2\cos^2 \!\!\varphi_{\!\scriptscriptstyle M}  \left( x + |M|^2\cos^2\!\! \varphi_{\!\scriptscriptstyle M}  \right)\,,$}\\
            I_\srm{E}(-p_\srm{E}^2) + I_\srm{R}(-p_\srm{E}^2)\,, & \text{otherwise}.
		 \end{cases}
\ee

As a concrete example, we consider $I(-p_\srm{E}^2)$ with $M = (500 + 50 i)~\text{MeV}$, in anticipation of the nonperturbative study in \sect{sec:num}, and $m = 800~\text{MeV}$, characteristic of the position of the first singularity of the gluon propagator~\cite{Cyrol:2018xeq,Binosi:2019ecz,Ferreira:2025tzo}.

\begin{figure}[!t]
    \centering
    \includegraphics[width = 0.49\textwidth]{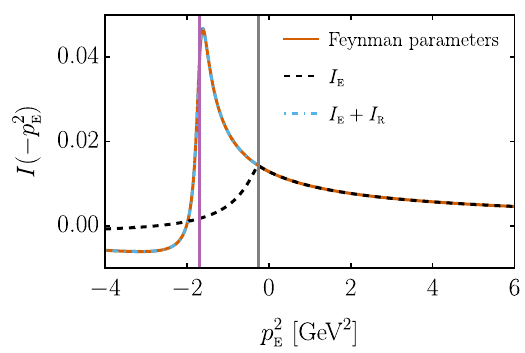}\hfil\includegraphics[width = 0.49\textwidth]{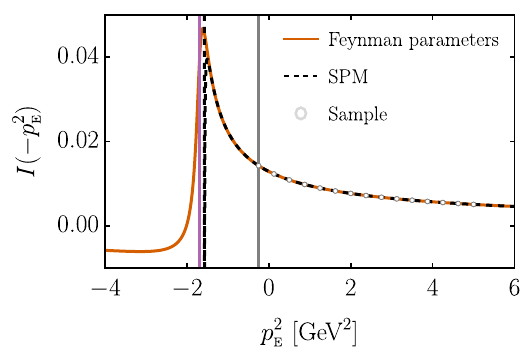}
    \caption{Left: $I(-p_\srm{E}^2)$ for $M = ( 500 + 50 i )$~MeV and $m = 800$~MeV, computed through Feynman parameters (orange continuous), \1eq{eq:Ires}, and the contribution from $I_\srm{E}(-p_\srm{E}^2)$ (black dashed). For $p_\srm{E}^2< - \textrm{Re}^2 (M)$, the residue term is added, $I_\srm{E}(-p_\srm{E}^2) + I_\srm{R}(-p_\srm{E}^2)$, to obtain the light blue dot-dashed line. Right: SPM extrapolation (black dashed) constructed from the 15 data points (circles) with $p_\srm{E}^2> - \textrm{Re}^2 (M)$, compared with the exact result $I(-p_\srm{E}^2)$ (orange). In both panels, $I_\srm{E}(-p_\srm{E}^2)$ is valid to the right of the gray vertical line [$p_\srm{E}^2>-\textrm{Re}^2(M)$], and the purple line marks the real part of the production threshold.}
    \label{fig:SPMex}
\end{figure}

On the left panel of \fig{fig:SPMex} we show the Feynman parametrization result of \1eq{eq:Ires} as an orange continuous line for real $p^2_\srm{E}$. In this case, the domain of validity of $I(-p_\srm{E}^2)$ becomes, from \1eq{eq:parab} with $p_\srm{E}^2 = x$ and $y = 0$,
\be\label{eq:parab_re}
p_\srm{E}^2 > - |M|^2\cos^2\!\! \varphi_{\!\scriptscriptstyle M}  = - \textrm{Re}^2(M) \,,
\ee
whose boundary is marked as a vertical gray line. Indeed, the contribution, $I_\srm{E}(-p^2)$, (black dashed line) is seen to match the Feynman parametrization result not only for space-like momenta, $p_\srm{E}^2>0$, but for $p_\srm{E}^2$ satisfying \1eq{eq:parab_re}. Instead, for $p_ \srm{E}^2<-\textrm{Re}^2(M)$, the Feynman parametrization result is reproduced by accounting for the residue term, \ie by the sum $I_\srm{E}(-p^2) + I_\srm{R}(-p^2)$ (light blue dot-dashed). Lastly, the vertical purple line marks the real part of the production threshold, \ie $-\textrm{Re}\left[(M + m)^2\right]$, which roughly coincides with the peak of $I(-p^2_\srm{E})$.

\subsection{SPM extrapolation}\label{subsec:bench}

In the nonperturbative case, the evaluation of the residue term is often impractical. In this case, an effective strategy for accessing the complex plane behavior beyond the domain of validity of the Euclidean integral is the SPM~\cite{Schlessinger:1968vsk, Tripolt:2018xeo, Binosi:2019ecz}. 
In what follows, we benchmark this method against \(I(-p_\srm{E}^2)\) to validate its application to the nonperturbative quark-gluon vertex.

We begin by selecting a sample of $n$ values of $I(-p^2_\srm{E})$, computed for real $p_\srm{E}^2>-\text{Re}^2(M)$, where the standard Euclidean integral is valid. Then, the SPM algorithm is used to construct a rational approximant that interpolates the sample data. A numerical analytic continuation is then obtained by simply complexifying the argument, $p_\srm{E}^2$, of the approximant.

On the right panel of \fig{fig:SPMex} we show $n = 15$ sample points (circles) and the SPM approximant constructed from them (black dashed curve). The latter is seen to correctly reproduce the exact $I(-p_\srm{E}^2)$  (orange continuous) until it approaches the production threshold (purple line). Near the threshold, the SPM develops a cluster of poles; the first one is denoted as the black dashed line becoming vertical. Past that point, the SPM output no longer reproduces the exact result, and is not shown.

\begin{figure}[!t]
\hspace*{-1.5cm}
    \includegraphics[scale=0.74]{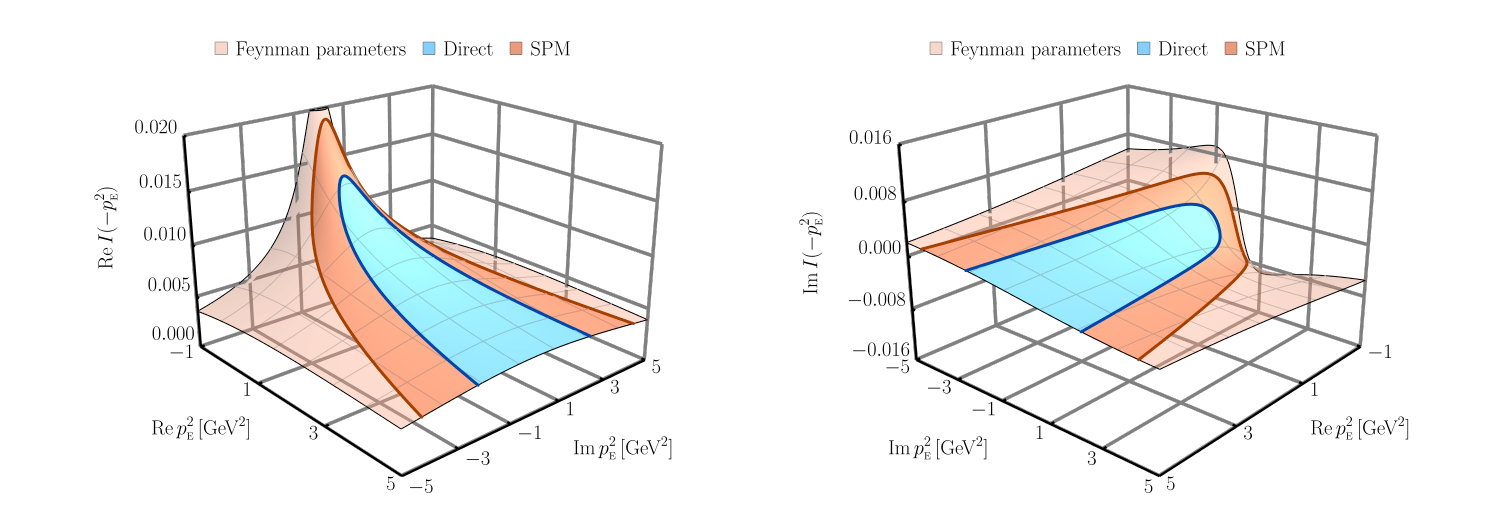}
\caption{Real (left) and imaginary (right) parts of $I(-p_\srm{E}^2)$ computed exactly through \1eq{eq:Ires} (pale orange) with $M = ( 500 + 50 i )$~MeV and $m = 800$~MeV. The domain of validity of the Euclidean integral is marked as light blue surfaces, while the orange surfaces show the SPM extrapolation.}
\label{fig:inttreeCCrescomplex}
\end{figure}

In \fig{fig:inttreeCCrescomplex} we show the results for complex $p_\srm{E}^2$. We find that the SPM accurately predicts the correct $I(-p_\srm{E}^2)$ in a region (marked in orange) significantly beyond the domain of validity of the Euclidean integral (light blue). In particular, the SPM is accurate in the parabolic domain
\be\label{eq:ext_parab}
y^2 \lessapprox 4 \textrm{Re}^2(M+m)  \left[ x + \textrm{Re}^2(M+m) \right] \,,
\ee
with apex near the square of the real part of the production threshold. We emphasize that the SPM correctly reproduces both the real and imaginary parts of $I(-p_\srm{E}^2)$, in spite of the fact that the sample data used to construct the approximant is purely real.

\section{Computations and results in the complex plane}\label{sec:num}

In this section we present the main results of the present work,
namely the structure of all eight vertex form factors $\lambda^{sg}_i(z)$ 
in an extended domain of the complex plane.
The key steps of our procedure may be 
summarized as follows: 

({\it i}) The starting point 
is the set of integrals 
determining the $\lambda^{sg}_i(p^2)$
in the Euclidean space, 
together with the accompanying 
structures, namely 
Eqs.(\ref{eq:lmbds})-(\ref{kernels}).

({\it ii}) 
For the quark propagator $S(p)$ entering 
in \1eq{eq:liSDE1} we use an {\it Ansatz} 
that contains complex-conjugate poles,
located at $p^2 =M^2$ and $p^2 =M^*{}^2$, 
to wit 
\be
S(p)=\frac{1}{2}\left[\frac{\slashed{p}+M}{p^2-M^2}+\frac{\slashed{p}+M^\ast}{p^2-M^*{}^2 }\right]\,.
\label{eq:quarkpoles}
\ee
The specific values employed in the 
numerical analysis, namely  
$M=(500+ 50 i)~\textrm{MeV}$ and 
$M^\ast =  (500-50i)~\textrm{MeV}$,
correspond to the results of  
\cite{Miramontes:2019mco}, obtained  
within the rainbow-ladder 
approximation. 
We emphasize that varying the 
values of $M$ within a range typical 
for this approximation 
does not change our findings qualitatively.

\begin{figure}[!t]
\centering
\includegraphics[width=\textwidth, keepaspectratio]{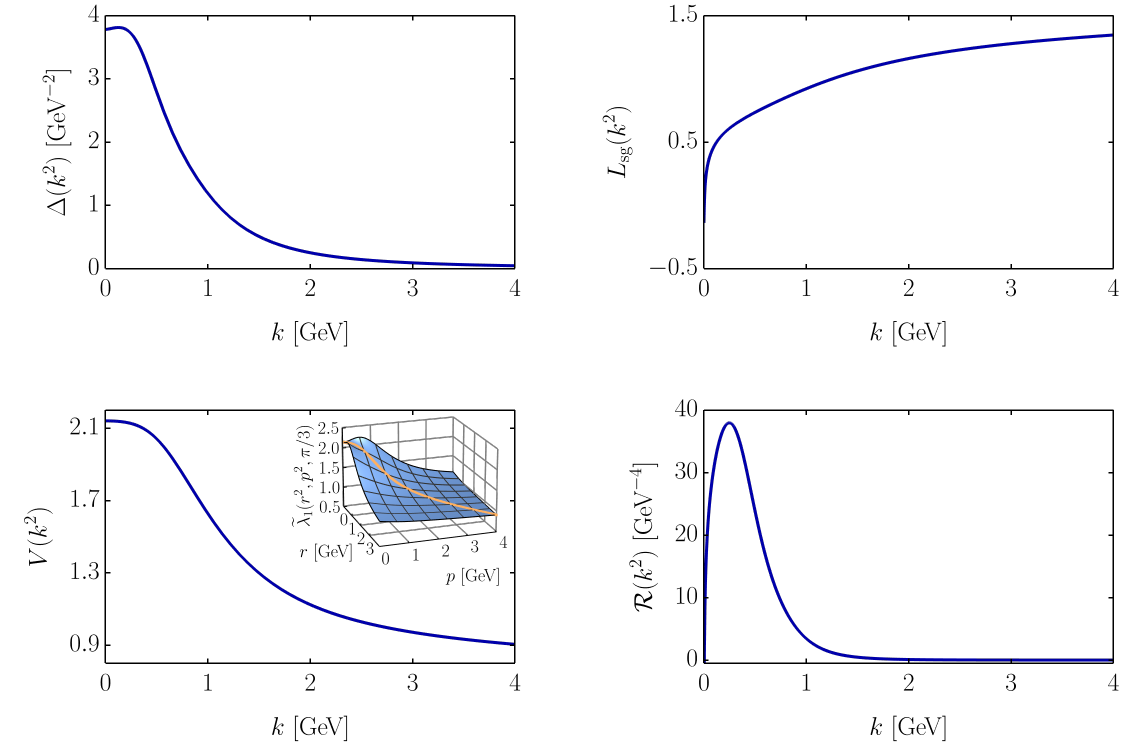}
\caption{The external input functions
$\Delta(k^2)$, $L_{sg}(k^2)$,  
$V(k^2)$, and ${\cal R}(k^2)$, 
discussed in 
item ({\it iii}).}
\label{fig:input}
\end{figure}

({\it iii}) The renormalization of these expressions is carried out by  
employing the version of the
momentum-subtraction 
(MOM) scheme~\cite{Celmaster:1979km,Hasenfratz:1980kn,Braaten:1981dv} known as \MOMt{}~\cite{Skullerud:2002ge,Kizilersu:2021jen,Aguilar:2023mam,Aguilar:2024ciu}. This scheme is defined using as reference 
precisely the soft-gluon limit of the quark-gluon vertex, 
namely through the condition 
 $\lambda_{1, \s R}^{sg}(\mu^2)=1$. 
Thus, one arrives at the renormalized 
set of integrals 
\be
\lambda_{i,\s{R}}^{sg}(p^2) = \lambda_{i,\s{Q}}^{sg}(p^2)+\left[1-\lambda_{i,\s{Q}}^{sg}(\mu^2)\right]\delta_{i1}\,,
\ee
where the index ``$R$" will be suppressed 
in what follows. 

({\it iv}) The ensuing numerical analysis requires the following inputs: 

\begin{figure}[!t]
    \hspace*{-1.5cm}
    \includegraphics[scale=0.74]{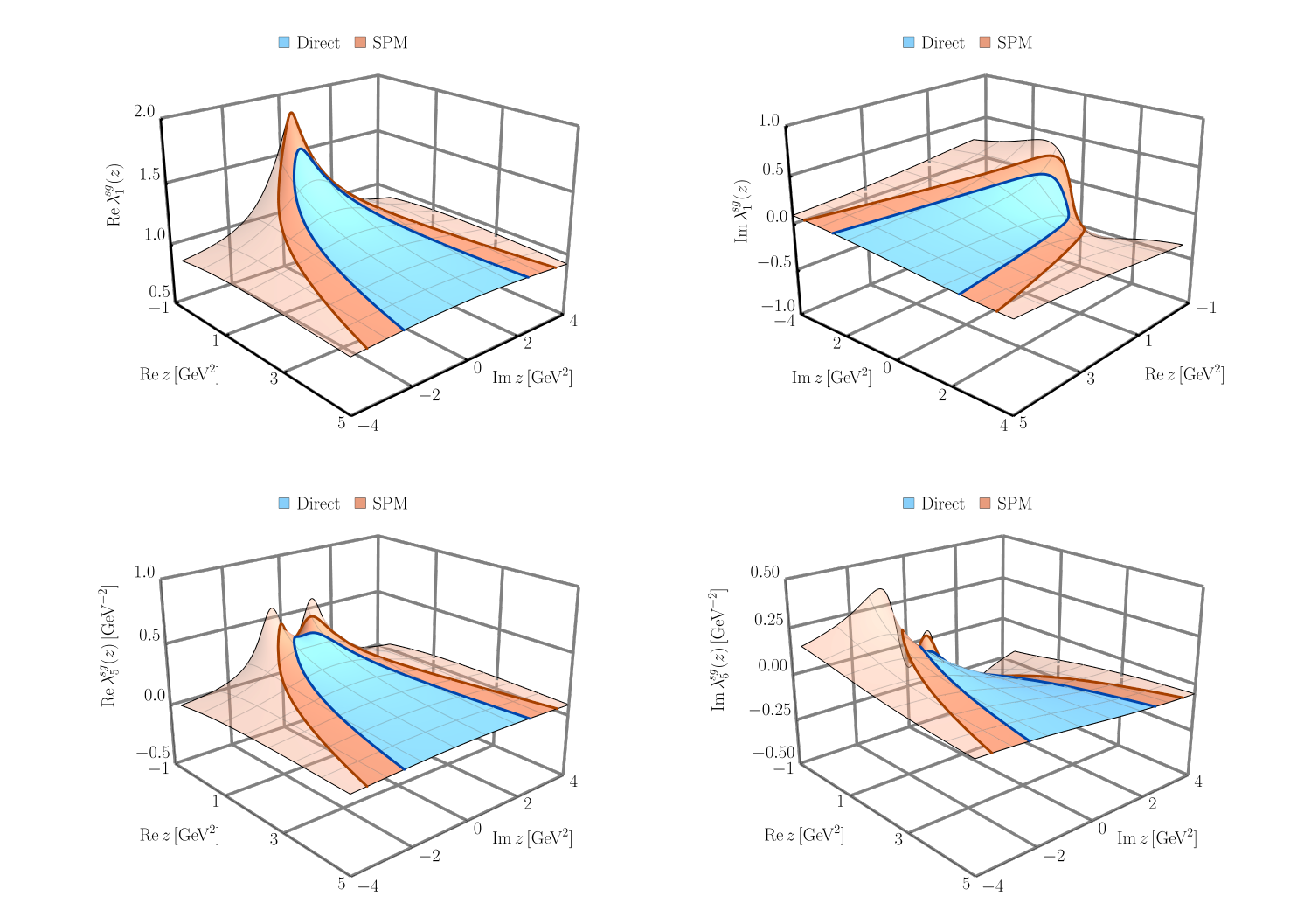}
    \caption{Real (left) and imaginary (right) parts of the chiral symmetry preserving form factors $\lambda_1^{sg}(z)$, and $\lambda_5^{sg}(z)$. Light blue surfaces mark the domain where direct Euclidean computations were performed, while orange surfaces show the region where the SPM extrapolation 
    is considered reliable.}
    \label{fig:L1}
\end{figure}

\begin{itemize} \item  For the gluon propagator 
we employ the fit to $N_f=2$ lattice data given in Eq.~(A1) and Tab.~II in \cite{Aguilar:2023mam}; it is shown in the upper-left panel of \fig{fig:input}, renormalized at $2~\textrm{GeV}$.

\begin{figure}[!t]
    \hspace*{-1.5cm}
    \includegraphics[scale=0.74]{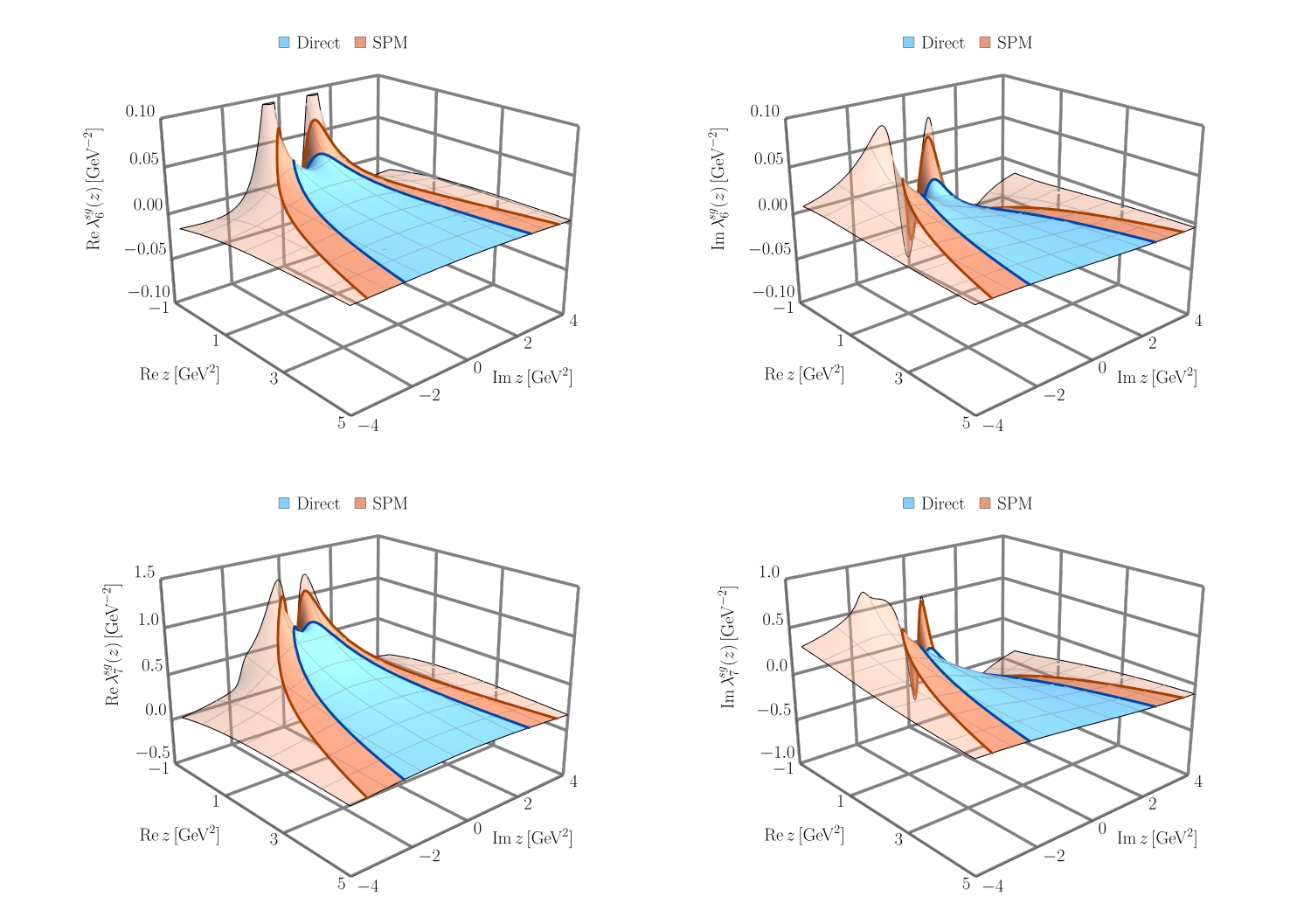}
    \caption{Real (left) and imaginary (right) parts of the chiral symmetry preserving form factors $\lambda_6^{sg}(z)$, and $\lambda_7^{sg}(z)$. Light blue surfaces mark the domain where direct Euclidean computations were performed, while orange surfaces show the region where the SPM extrapolation is considered reliable.}
    \label{fig:LSymmetric}
\end{figure}

\item  
For the form factor $L_{sg}$, 
shown in the upper-right panel of \fig{fig:input}, 
we use the fit to lattice data given in Eq.~(A1) and Tab.~II in \cite{Aguilar:2023mam}. 

\item   
For the function $V(q^2)$, shown in the 
lower-left panel of \fig{fig:input}, 
we use the parametrization given in Eqs.~(6.5),~(6.6) and Tab.~I of \cite{Ferreira:2025wpu}.
As explained in \sect{sec:intro} and elaborated in \cite{Ferreira:2025wpu, Ferreira:2026gbe}, 
$V(q^2)$ is identified with 
the symmetric slice, $q^2=p^2=r^2$, of the classical form factor in the quark-gluon vertex, $\widetilde{\lambda}_1$, shown in the inset. 

\item   
The above quantities 
are combined to form 
the function 
\mbox{${\cal R}(k^2)=V^2(k^2)\Delta^2(k^2)L_{sg}(k^2)$}, introduced in 
\1eq{eq:calRdef}; the resulting curve is shown in the lower-right panel of \fig{fig:input}.

\end{itemize} 

({\it v}) Then, the external momentum 
$p^2$ entering in Eqs.(\ref{eq:lmbds})-(\ref{kernels})
is 
complexified, by setting 
\be
p^2 \to z \,, \qquad z\in\mathbb{C}\,,
\ee
while the integration (loop) momentum $k$ remains real and positive.

\begin{figure}[!t]
    \hspace*{-1.5cm}
    \includegraphics[scale=0.74]{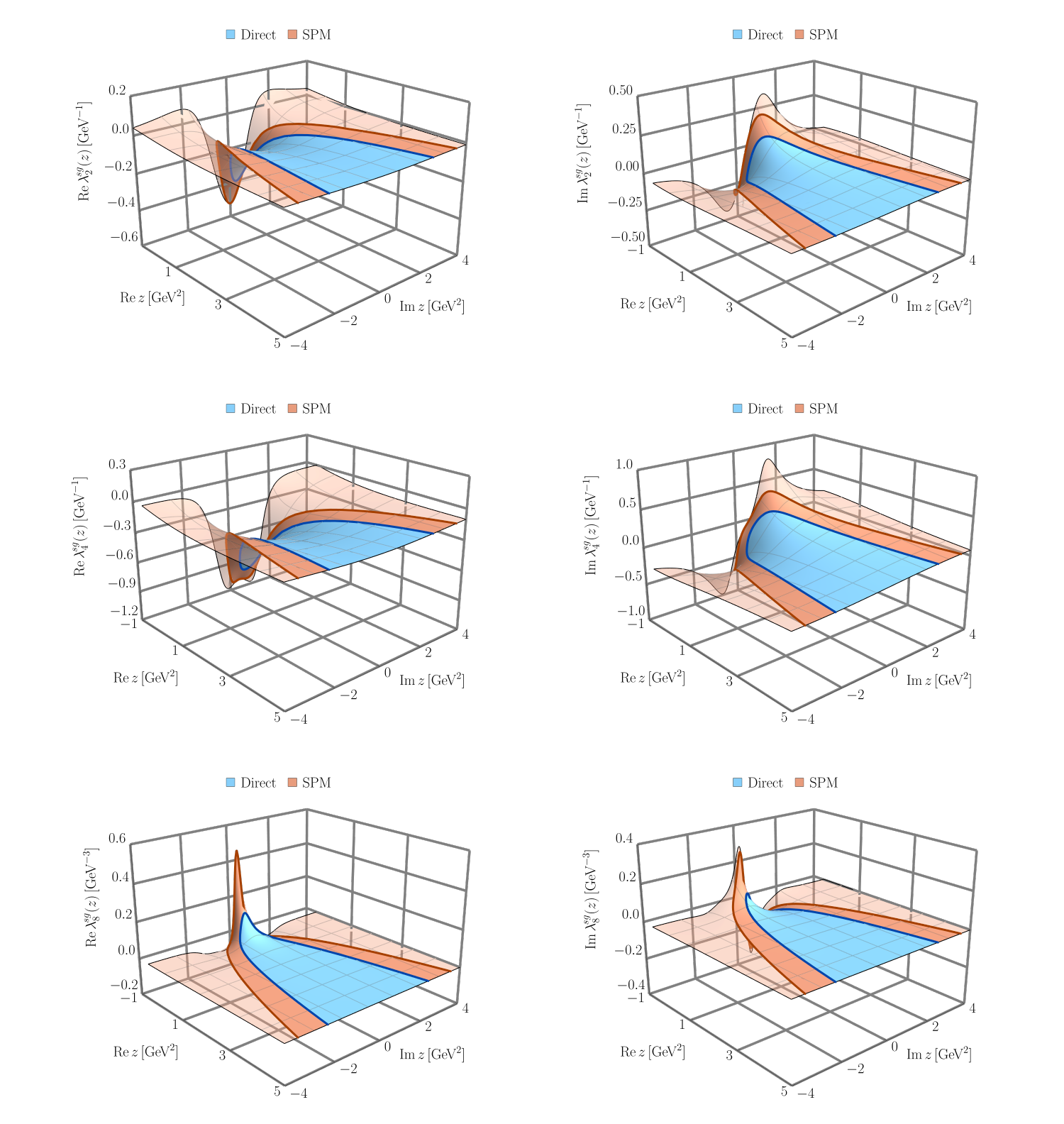}
    \caption{Real (left) and imaginary (right) parts of the chiral symmetry breaking form factors $\lambda_2^{sg}(z)$, $\lambda_4^{sg}(z)$, and $\lambda_8^{sg}(z)$. Light blue surfaces mark the domain where direct Euclidean computations were performed, while orange surfaces show the region where the SPM extrapolation is considered reliable.}
    \label{fig:Lbreaking}
\end{figure}

({\it vi})
The computation of the  
vertex integrals within the maximal accessible region, given by \1eq{eq:parab}, 
yields the $\lambda^{sg}_i(z) = {\rm R e}\,\lambda^{sg}_i(z) + i \,{\rm I m} \,\lambda^{sg}_i(z) $. The results for 
$\lambda^{sg}_1(z)$ and $\lambda^{sg}_5(z)$
are 
shown in \fig{fig:L1}, 
for $\lambda^{sg}_6(z)$ 
and $\lambda^{sg}_7(z)$
in
\fig{fig:LSymmetric}, and for
$\lambda^{sg}_2(z)$, $\lambda^{sg}_4(z)$, 
and $\lambda^{sg}_8(z)$
in \fig{fig:Lbreaking}. 
The accessible domains 
are indicated by the light blue surfaces; they  
are bounded by the dark blue parabolas marked in these figures, and are labeled ``direct''. 
 Using \1eq{eq:parab}, with $x=\textrm{Re}\,z$ and $y=\textrm{Im}\,z$, 
 the parabola obtained from the 
 direct computation 
is given by 
\be
y^2= x\,\alpha_{\s{\textrm{direct}}} \,+\beta_{\s{\textrm{direct}}}\,,\quad\textrm{with}\quad\alpha_{\s{\textrm{direct}}}=1.00~\textrm{GeV}^2\,,\qquad\beta_{\s{\textrm{direct}}}=0.25~\textrm{GeV}^4\,.
\ee

({\it vii}) Once the form factors $\lambda_i^{sg}(z)$ have been computed within the maximal accessible region, we investigate their SPM extrapolation \cite{Schlessinger:1968vsk, Tripolt:2018xeo, Binosi:2019ecz}, through the methods described in \sect{sec:anstruct}. In particular, for each form factor we use $n=15$ randomly selected points from the available datasets to construct the corresponding SPM approximant, whose coefficients are listed in \tab{tab:coeffsspm} of \appref{app:coefs}. 
We stress that, although a relatively small sample size is employed, the robustness of the extrapolations has been validated against constructions using significantly larger input-data sets. The chosen configuration captures the essential features of these more elaborate cases, while offering a simpler setup.
In addition, in all cases, the resulting SPM approximants reproduce very accurately the 
data obtained from the direct calculation 
(\ie blue regions in  \figs3{fig:L1}{fig:LSymmetric}{fig:Lbreaking}).

({\it viii}) 
The red parabolas shown in \figs3{fig:L1}{fig:LSymmetric}{fig:Lbreaking} delineate the 
orange regions where the SPM-derived form factors are expected to be reliable. The extend 
of these regions is limited by the 
position of the first singularity in the
$\lambda_1(z)$ so obtained.

({\it ix}) Since SPM extrapolations constructed from different data subsets may exhibit slightly different analytic structures, we generate $500$ SPM interpolants for $\lambda_1^{sg}(z)$ and extract the position of the first singularity, $z_{\s{\textrm{s}}}=-0.94~\textrm{GeV}^2$, along with its standard deviation, \mbox{$\sigma_{\s{\textrm{s}}}=0.20~\textrm{GeV}^2$}. To ensure numerical stability, calculations should avoid this singularity; we therefore define the apex of the red parabola to lie at a distance of $2\sigma_{\s{\textrm{s}}}$ away from $z_{\s{\textrm{s}}}$. The extended parabola is characterized by
\begin{align}
\alpha_{\s{\textrm{SPM}}}&=2.19~\textrm{GeV}^2\,,&\beta_{\s{\textrm{SPM}}}&=1.20~\textrm{GeV}^4\,.
\end{align}
Notably, the apexes, $x^{\s{\textrm{apex}}}=-\beta/\alpha$, of these two parabolic regions are located at
\begin{align}
x_{\s{\textrm{direct}}}^{\s{\textrm{apex}}}&=-0.25~\textrm{GeV}^2 \,,&
x_{\s{\textrm{SPM}}}^{\s{\textrm{apex}}}=&-0.54~\textrm{GeV}^2\,;
\end{align}
thus the SPM increases our reach into the time-like region 
by a factor of $2.16$. 

\section{Discussion and Conclusions}\label{sec:conc}

In the present work we have explored 
for the first time the structure of the quark-gluon vertex in the complex plane. We have used a version of the vertex that is compatible with the recently developed symmetry-preserving 
approach for the study of  
meson properties beyond the rainbow-ladder approximation \cite{Miramontes:2025imd,Ferreira:2025wpu,Ferreira:2026gbe}. 
The analysis presented has focused on the so-called ``soft-gluon" limit, which 
reduces the momentum-dependence of the 
corresponding vertex form factors 
$\lambda^{sg}_i(p^2)$
to a single momentum variable.

The complexification of this variable, 
$p^2 \to z \,, z\in\mathbb{C}$, 
inside the integrals that 
define the form factors, yields the complex functions $\lambda^{sg}_i(z)$ within a concrete domain of the variable $z$, delimited by a characteristic parabola. 
The extent of this complex domain is 
determined by the appearance of the first singularity in the integrands of the vertex integrals, 
as has been illustrated in detail 
by means of an exactly solvable toy model (triangle diagram).
This domain may be extended considerably 
by resorting to the SPM, which allows 
the extrapolation of $\lambda^{sg}_i(z)$
up until the appearance of  
Landau singularities, \eg branch cuts,
associated with the onset of physical 
processes (particle production thresholds). 

The analysis presented sets the stage for the 
exploration of the complex structure of the quark-gluon vertex beyond the soft-gluon limit approximation. 
In the most general case, the form factors $\lambda_i(q,r,-p)$ may be parametrized in terms of the squares of two momenta, say $q^2$ and $p^2$, and the angle between them~~\cite{Huber:2022nzs,Huber:2023uzd}, which remains real even if the squared momenta are complexified. Evidently, the complexification of the gluon momentum squared, $q^2$, presents the greatest challenge, since in this case the SDE for the $\lambda_i(q,r,-p)$ will inevitably involve the complex plane behavior of the gauge sector Green's functions, $\Delta(q^2)$ and $\Ls(q^2)$. However, in the study of light mesons, it is possible to channel the momenta in the vertex SDE, gap equation, and BSE, such that all of the gluon momenta appearing in these equations remain space-like. In this case, the coupled system of equations can be solved self-consistently for $\lambda_i(q,r,-p)$ and $S(p)$, through the standard Euclidean integrals and the known space-like forms of $\Delta(q^2)$ and $\Ls(q^2)$, with $p^2$ in the same parabolic domain of \1eq{eq:parab}.

Throughout this work, we have reduced the 
complexity of the problem by resorting 
to the {\it Ansatz} of \1eq{eq:quarkpoles}
for the  quark-propagator $S(p)$ entering in the 
diagrams that define the $\lambda^{sg}_i(z)$. 
The next important step in this quest is to consider instead a quark-propagator that is determined from the dynamical equation that controls its evolution. 
Specifically, the analysis must be extended to 
the {\it coupled system} of dynamical equation composed 
by the integral expressions in \1eq{eq:lmbds}, furnishing the $\lambda^{sg}_i(z)$,
and the gap equation that determines $S(p)$ \cite{Eichmann:2016yit, Sanchis-Alepuz:2017jjd, Huber:2025cbd,Eichmann:2025wgs}. 
Such an analysis is expected 
to provide further key insights on the presence or absence of complex-conjugate poles 
in $S(p)$, especially in view of the studies 
presented in \cite{Pawlowski:2024kxc} and 
\cite{Wieland:2026iml}.

The present analysis, together with the 
complementary directions mentioned above, 
pave the way towards a comprehensive  
study of the meson masses from the BSE 
derived in \cite{Miramontes:2025imd,Ferreira:2025wpu}. 
In particular, the knowledge of the complex structure 
of both the quark-propagator and the quark-gluon vertex 
makes possible the extension of the preliminary study 
presented in \cite{Ferreira:2026gbe} to include further mesonic states.
In particular, ``on-shell" computations 
of the BSE
may access heavier states, while 
SPM extrapolations may be carried out more 
reliably and with reduced error. 
We hope to present results in these directions 
in the near future. 

\section*{Acknowledgments}\label{sec:Acknow}
A.S.M., J.M.M. and J.P. are funded by the Spanish MICINN grants PID2020-113334GB-I00 and PID2023-151418NB-I00, the Generalitat Valenciana grant CIPROM/2022/66, and CEX2023-001292-S by MCIU/AEI. Part of the 
computations have been carried out at the CEAFMC and Universidad de Huelva High Performance Computer (HPC@UHU), funded by FEDER/MINECO project UNHU-15CE-2848. The authors thank R. Alkofer and A. Santamaria for useful discussions.

\clearpage

\appendix

\section{SPM coefficients}\label{app:coefs}

In this appendix we list the coefficients defining the SPM approximants constructed in \sect{sec:num} for each form factor $\lambda_i^{sg}(z)$.

\renewcommand{\arraystretch}{0.8} 
\begin{table}[!h]
\begin{tabular}{c|c c c c c c c c}
\toprule\toprule
 & $z^{0}$ & $z^{1}$ & $z^{2}$ & $z^{3}$ & $z^{4}$ & $z^{5}$ & $z^{6}$ & $z^{7}$ \\
\midrule
\multicolumn{9}{c}{$\lambda_1^{sg}$}\\
\midrule
$P(z)$ & $1234.49$ & $120.096$ & $-476.363$ & $1236.88$ & $157.774$ & \
$-134.857$ & $0.569072$ & $0.722294$ \\
\hline
$Q(z)$ & $834.481$ & $321.269$ & $-399.175$ & $790.723$ & $348.185$ & \
$-146.87$ & $-2.86829$ & $1$ \\
\midrule\midrule
\multicolumn{9}{c}{$\lambda_2^{sg}$}\\
\midrule
$P(z)$ & $203.983$ & $-281.501$ & $160.913$ & $-49.7817$ & $8.10392$ \
& $-0.449153$ & $-0.0301992$ & $-0.000368969$ \\
\hline
$Q(z)$ & $-721.44$ & $273.08$ & $215.144$ & $-91.6116$ & $-32.2671$ & \
$31.9479$ & $-9.09329$ & $1$ \\
\midrule\midrule
\multicolumn{9}{c}{$\lambda_4^{sg}$}\\
\midrule
$P(z)$ & $401.724$ & $-384.066$ & $81.7961$ & $61.5987$ & $-42.0921$ \
& $9.72442$ & $-0.801812$ & $-0.00163241$ \\
\hline
$Q(z)$ & $-530.156$ & $190.627$ & $108.537$ & $-19.903$ & $-74.2213$ \
& $48.1652$ & $-11.5555$ & $1$ \\
\midrule\midrule
\multicolumn{9}{c}{$\lambda_5^{sg}$}\\
\midrule
$P(z)$ & $235.752$ & $87.9066$ & $-155.534$ & $31.4565$ & $15.0464$ & \
$-6.53072$ & $0.743032$ & $0.00193203$ \\
\hline
$Q(z)$ & $622.694$ & $362.71$ & $-160.196$ & $-145.916$ & $73.0643$ & \
$6.10274$ & $-7.06348$ & $1$ \\
\midrule\midrule
\multicolumn{9}{c}{$\lambda_6^{sg}$}\\
\midrule
$P(z)$ & $16.0502$ & $21.6751$ & $-35.8044$ & $17.1213$ & $-3.36948$ \
& $0.20401$ & $0.00977832$ & $0.0000306083$ \\
\hline
$Q(z)$ & $560.744$ & $307.912$ & $-160.898$ & $-166.377$ & $85.6607$ \
& $4.56875$ & $-7.36035$ & $1$ \\
\midrule\midrule
\multicolumn{9}{c}{$\lambda_7^{sg}$}\\
\midrule
$P(z)$ & $534.01$ & $813.058$ & $-166.895$ & $48.9318$ & $-77.6561$ & \
$21.3911$ & $-1.47126$ & $-0.00388936$ \\
\hline
$Q(z)$ & $712.48$ & $1180.85$ & $336.807$ & $-53.6837$ & $-50.5117$ & \
$-28.0326$ & $12.524$ & $-1$ \\
\midrule\midrule
\multicolumn{9}{c}{$\lambda_8^{sg}$}\\
\midrule
$P(z)$ & $8.76334$ & $-3.53096$ & $-1.20631$ & $0.459484$ & \
$0.0643481$ & $-0.0266072$ & $0.00132844$ & $-0.0000412388$ \\
\hline
$Q(z)$ & $114.724$ & $214.929$ & $53.1718$ & $-61.2133$ & $-20.1227$ \
& $-1.21792$ & $5.8499$ & $-1$ \\
\bottomrule\bottomrule
\end{tabular}
\caption{Coefficients of the SPM approximants for each form factor, $\lambda_i^{sg}(z)$. The approximants were built as $P(z)/Q(z)$, \1eq{eq:spmdef}, and were constructed using $15$ data points (see \sect{sec:num}).}
\label{tab:coeffsspm}
\end{table}

For convenience, we express the SPM approximant in rational form, \ie
\be\label{eq:spmdef}
\lambda_i^\srm{SPM}(z) = \frac{P^k_i(z)}{Q^\ell_i(z)}\,,\qquad [k\,,\ell] = \begin{cases}
[n/2-1,\,n/2]\,,  & \text{for $n$ even},  \\
[(n-1)/2,\,(n-1)/2]\,,& \text{for $n$ odd},
\end{cases}
\ee
where $P^m_i$ and $Q^m_i$ are polynomials of degree $m$, rather than the continued fraction used internally in the algorithm~\cite{Schlessinger:1968vsk, Tripolt:2018xeo, Binosi:2019ecz}. 

\bibliography{bibliography.bib}

\end{document}